\documentclass[aps,pra,superscriptaddress,twocolumn,showpacs,titlepage=false]{revtex4}
\usepackage{graphicx}
\usepackage{amsmath}
\usepackage{amssymb}
\usepackage{subfigure}
\usepackage{natbib}
\usepackage{color}
\usepackage[breaklinks=true,colorlinks=true,linkcolor=blue,urlcolor=blue,citecolor=blue]{hyperref}
\def\bra#1{{\left\langle #1 \right|}}
\def\ket#1{{\left| #1 \right\rangle}}
\def\braket#1{{\left\langle #1 \right\rangle}}

\usepackage{xcolor}	 
\definecolor{jesusgreen}{RGB}{0,127,0}

\usepackage[T1]{fontenc}
\usepackage{tabularx}
\usepackage{graphicx}
\usepackage{amsmath}
\usepackage{amsfonts}
\usepackage{verbatim}
\usepackage{bbm, bm}
\usepackage{amssymb}
\usepackage{multirow}
\usepackage{enumitem}
\usepackage{subfigure}

\begin{document}

\title{Designing quantum experiments with a genetic algorithm}
\author{Rosanna Nichols}
\affiliation{Centre for the Mathematics and Theoretical Physics of Quantum Non-Equilibrium Systems (CQNE), School of Mathematical Sciences, University of Nottingham, University Park, Nottingham, NG7 2RD, UK.}
\author{Lana Mineh}
\affiliation{Quantum Engineering Centre for Doctoral Training, University of Bristol, Bristol, UK.}
\affiliation{Now at: School of Mathematics, University of Bristol, UK.}
\author{Jes{\'u}s Rubio}
\affiliation{Department of Physics and Astronomy, University of Sussex, Brighton, BN1 9QH, UK.}
\author{Jonathan C. F. Matthews}
\affiliation{Quantum Engineering Technology Labs, H. H. Wills Physics Laboratory and Department of Electrical \& Electronic Engineering, University of Bristol, BS8 1FD, UK.}
\author{Paul A. Knott}
\thanks{Contact email address: Paul.Knott@Nottingham.ac.uk}
\affiliation{Centre for the Mathematics and Theoretical Physics of Quantum Non-Equilibrium Systems (CQNE), School of Mathematical Sciences, University of Nottingham, University Park, Nottingham, NG7 2RD, UK.}
\date{\today}

\begin{abstract}
We introduce a genetic algorithm that designs quantum optics experiments for engineering quantum states with specific properties. Our algorithm is powerful and flexible, and can easily be modified to find methods of engineering states for a range of applications. Here we focus on quantum metrology. First, we consider the noise-free case, and use the algorithm to find quantum states with a large quantum Fisher information (QFI). We find methods, which only involve experimental elements that are available with current or near-future technology, for engineering quantum states with up to a 100-fold improvement over the best classical state, and a 20-fold improvement over the optimal Gaussian state. Such states are a superposition of the vacuum with a large number of photons (around $80$), and can hence be seen as Schr\"odinger-cat-like states. We then apply the two most dominant noise sources in our setting -- photon loss and imperfect heralding -- and use the algorithm to find quantum states that still improve over the optimal Gaussian state with realistic levels of noise. This will open up experimental and technological work in using exotic non-Gaussian states for quantum-enhanced phase measurements. Finally, we use the Bayesian mean square error to look beyond the regime of validity of the QFI, finding quantum states with precision enhancements over the alternatives even when the experiment operates in the regime of limited data.
\end{abstract}
\maketitle

Engineering quantum states with specific properties plays a part in all experiments and technologies in quantum physics. But designing optimal experiments to engineer such states can be challenging, in part due to the counter-intuitive nature of the theory. This has prompted a number of recent works allocating the task of experiment-design to artificial intelligence and machine learning \cite{knott2016search,krenn2016automated,melnikov2018active,arrazola2018machine,sabapathy2018near}. Many of the computer-designed experiments have surpassed the best human-designed alternatives.

Building on this approach, we introduce a genetic algorithm, named AdaQuantum \footnote{The algorithm, AdaQuantum, is named after Ada Lovelace, the world's first computer programmer, and resident of Nottingham, where our own algorithm was born.}, that designs quantum optics experiments to produce quantum states with the required properties. The algorithm is flexible and modifiable, so that researchers with a range of requirements and available quantum-optics equipment can use it to design and optimise their experiments, as well as facilitating theoretical research in quantum state engineering and experimental design. In addition, the algorithm has a powerful computational engine, so that it efficiently and effectively searches for the optimal experimental designs in a given setting, and can allow for truncations of the Hilbert space up to $170$ photons, enabling even more exotic states to be found. A selection of the experiments designed by AdaQuantum are shown schematically in Fig.~\ref{example_schematics}.

\begin{figure}[]
	\centering
	\includegraphics[trim=7.4cm 8cm 4cm 6cm, clip=true, width=1.02\linewidth]{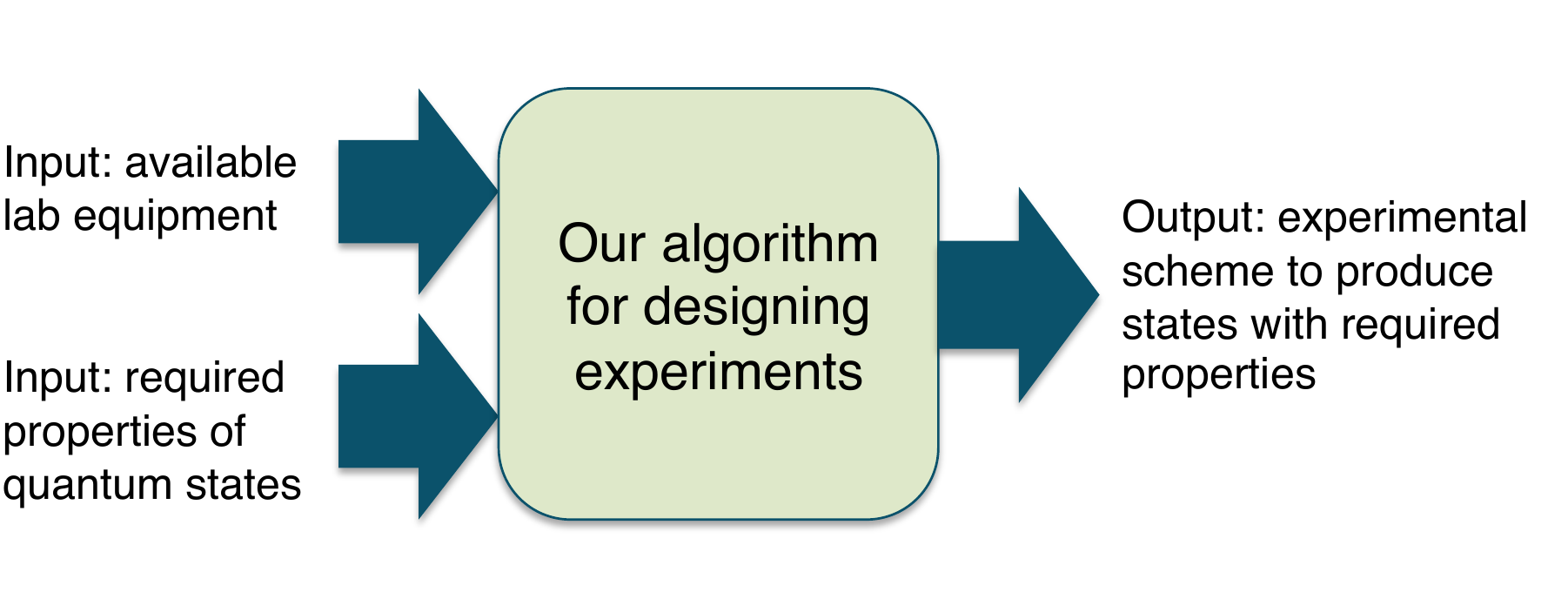}
	\caption{A flow chart illustrating the overall structure and usage of our algorithm, AdaQuantum, for designing quantum experiments.}
	\label{fig:flowchart_overall_structure}
\end{figure}

To run AdaQuantum we must specify two things: first, we specify the \emph{toolbox} of equipment that is available to construct our experiment, or more specifically, which quantum states, operations, and measurements we would like AdaQuantum to optimise over. Second, we must specify a fitness function, which allows us to quantify \emph{how good} the quantum state outputted by AdaQuantum is for our purposes. In principle, any fitness function that takes as input a quantum state, and outputs a real number, can be used. Given this, AdaQuantum then performs an automated search to find arrangements and parameter settings of the experiential equipment that produce a state that maximises (or minimises) the fitness function. The flexibility built into AdaQuantum means that different fitness functions can easily be implemented, such as the fidelity with useful quantum states such as Schr\"odinger-cat states or states used in optical quantum computation \cite{gottesman2001encoding}; or measures of entanglement, coherence, non-classicality, and so on. A flow chart illustrating the overall structure and usage of our algorithm is shown in Fig.~\ref{fig:flowchart_overall_structure}.
%

As a demonstration of the great potential of our algorithm, in this work we focus on fitness functions relevant to the field of quantum metrology, and we will see how its application reveals important findings that will help to design better metrological protocols. We first consider noise-free experiments, and use as our fitness function the quantum Fisher information (QFI) of the (pure) output state. The QFI allows us to quantify how precisely a given state can measure a phase shift in an interferometer -- a valuable fitness function for optical quantum metrology \cite{braunstein1994statistical,caves1981quantum}. Next, we apply the two most dominant noise sources in our quantum optics setting -- photon loss at the output, and imperfect heralding measurements -- and use the mixed state QFI as our fitness function.

The third fitness function that we optimise is the Bayesian mean square error (BMSE). Although the QFI plays a crucial role in quantum metrology, using the theory associated with it requires a number of assumptions that are not always fulfilled in experiments. For example, it often assumes that a large number of repetitions has taken place and that certain prior information was available \cite{jesus2017, berry2012infinite, rafal2015, haase2018jul}. The BMSE, on the other hand, factors in and allows us to control both of these explicitly, and thus gives a reliable measure of the phase-measuring capability of a given state in realistic experimental settings \cite{jesus2018}.

Our main results obtained by AdaQuantum are threefold: First, in the noiseless case we find a number of methods of producing quantum states with QFIs up to $20$ times larger than the optimal Gaussian state, which amounts to a $100$-fold improvement over the best classical state. We show that these states can be thought of as Schr\"odinger-cat-like states, as they contain superpositions of the vacuum with a large numbers of photons (around $80$). Second, we find methods of producing states that can still beat the optimal Gaussian state, even when realistic levels of the dominant noise sources (photon loss in the heralding measurements and in the final state) are included. Finally, using the BMSE for the problem of measuring a phase shift in an interferometer, we find states that beat both the optimal Gaussian state and the coherent state when the experiment operates with realistic prior knowledge and a reasonable number of experimental repetitions. This will enable experiments to create exotic states with enhanced phase-measuring capabilities. \\

\section{Using a genetic algorithm to design experiments}
\label{sec:usingageneticalgorithmtodesignexperiments}

\subsection{Optical quantum state engineering}
\label{sec:experimentalquantumoptics}

\begin{figure}[]
	\centering
	\includegraphics[width=1\linewidth]{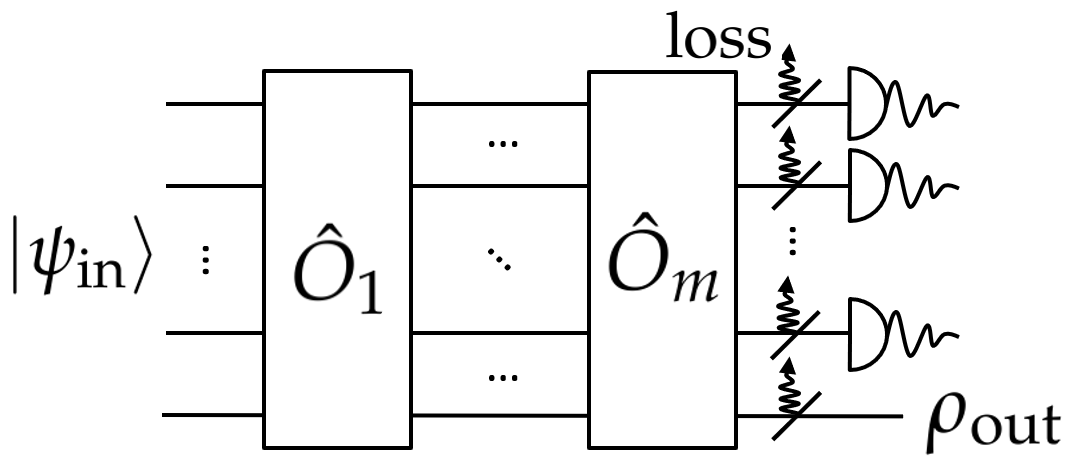}
	\caption{The generic blueprint for the quantum optics experiments that we use an algorithm to optimise. An input state, $ \ket{\psi}_\mathrm{in} $, is acted on by a series of operators, $\hat{O}_i$, before heralding measurements are performed to produce an output state, $\rho_{\mathrm{out}}$.}
	\label{fig:stateengineeringscheme}
\end{figure}

The quantum optics experiments we design follow the blueprint depicted in Fig.~\ref{fig:stateengineeringscheme}. In this scheme, we start with an $N$-mode state, $ \ket{\psi}_\mathrm{in} $, which consists of independent one- and two-mode states. This state is acted on by up to $m$ operators in sequence, which each act on one or two modes, with the two-mode operations serving to mix and entangle the modes. Finally, in $ N - 1 $ of the output modes heralding measurements are performed, modelled by POVMs. When each of these heralding measurements are simultaneously successful, the state in the $N^{\mathrm{th}}$ mode is the output state, which we then apply our fitness function to. With appropriate choices of experimental elements (which we introduce below), many of the state engineering schemes in the literature fit into our blueprint illustrated in Fig.~\ref{fig:stateengineeringscheme} \cite{gerrits2010generation, bartley2012multiphoton, ourjoumtsev2007generation, huang2015optical, etesse2015experimental}.

Although $N$ is flexible, here we concentrate on $ N = 2 $ modes because having more modes is more difficult for both experiments and for our computer simulations (later we outline future work to overcome the computational challenges and enable $ N > 2 $ modes). Experimentally, having more modes substantially lowers the overall probability for producing the desired output state, and adds significant additional noise.

To design an experiment using the blueprint in Fig.~\ref{fig:stateengineeringscheme}, we need a toolbox of states, operations and measurements -- the full list of these elements that we consider is given in Table~\ref{fig:toolbox}. Our focus is on producing experimental designs that are feasible with current or near future technology, and hence our toolbox only contains such elements. Other elements can easily be added to this list, and the list can be customised to match the capabilities of a particular laboratory.


%
\begin{table}[t]
{\renewcommand{\arraystretch}{1.5} 
\begin{tabular}{l|l}	
			States~ & $ \ket{n} $, $ \ket{\alpha} $, $ \ket{\zeta}_i $, $ \ket{\zeta}_{ij} $ \\
			\hline
			Operations~ & $ \hat{D}(\alpha) $, $ \hat{S}_i(\zeta) $, $ \hat{S}_{ij}(\zeta) $, $ \exp{(i \hat{n} \phi)} $, $ \hat{U}_{ij}(T) $ \\
			\hline
			Measurements~ & $ \ket{n}\bra{n} $, Bucket, Multiplex, $ \ket{x_\lambda }\bra{x_\lambda} $
\end{tabular}}	
	\caption{The states, operations and measurements that make up our toolbox of experimentally feasible elements. Here, $i$ and $j$ are arbitrary modes. The identity operator is also included in all runs of the algorithm.}
	\label{fig:toolbox}
\end{table}


Here we only introduce the most important details of the toolbox; more details can be found in Appendix \ref{APX:experimentalquantumoptics}. Firstly, the input states we include are the single-mode squeezed vacuum $ \ket{\zeta}_i$ the two-mode squeezed vacuum (TMSV) $|\zeta\rangle_{ij}$, the coherent state $|\alpha\rangle$, and Fock states $|n\rangle$; the parameters $z$, $\alpha$ and $n$ are constrained by what is possible experimentally.

Next are the operators, of which the most important is the beam splitter $\hat{U}_{ij}(T)$, where $T$ is the probability of transmission, which serves to mix and entangle the two modes, enabling more exotic and useful states to be produced when part of the entangled state is measured. The other operators we use are the displacement operator $\hat{D}(\alpha)$, the phase shift $\exp{(i\hat{n}\phi)}$, and the single- and two- mode squeezing operators, given by $ \hat{S}_i(\zeta) $ and $ \hat{S}_{ij}(\zeta) $ respectively.

The first measurement we include is the \emph{photon number resolving detection} (PNRD), given by $\ket{n}\bra{n}$. However, due to the difficulty of implementing a PNRD we include two easier measurements, the \emph{bucket detector} \cite{Cova96APDs} and the \emph{multiplex detector} \cite{matthews2016towards}. The latter is achieved by separating the mode you want to measure into several modes via spatial- or time-multiplexing \cite{xiang2011entanglement, matthews2016towards, achilles2004photon}, followed by a bucket detection performed on each of these separated modes. Finally we include the homodyne detector, which is the projection onto a line in phase space, characterised by the projector $\ket{x_\lambda}\bra{x_\lambda}$.

We focus on the two most dominant noise sources in optics experiments taking the form of Fig.~\ref{fig:stateengineeringscheme}: i) imperfect photon detectors, which can be modelled as photon loss prior to detection, and ii) photon loss on the output state. Noise is also present in the initial state preparation and in implementing the operations, but we do not include these as they are typically smaller than the detection and output-state loss (though future work will also incorporate these). See Appendix \ref{APX:noise} for details of how we simulate noise.

\subsection{Our genetic algorithm}

We now introduce and describe our algorithm, AdaQuantum, that designs quantum optics experiments. To run our algorithm, the user first has to specify which input states, operators and measurements they would like the algorithm to search over. This is done via the user interface shown in Fig.~\ref{fig:UI - toolbox}. As well as selecting the toolbox, the user can specify the ranges of the parameters, the loss rates of the detectors, the loss rate on the final state, and a number of other details that we will introduce below.

Once a user has specified their toolbox, they next tell the algorithm what kind of quantum state they require by specifying a fitness function. The fitness function must take as input a quantum state, and output a real number that we wish to maximise (or minimise). In section \ref{sec:results} we introduce three different fitness functions that quantify the performance of the state for measuring a phase shift, but in principle any fitness function that can be written in the above-mentioned form can be incorporated into the algorithm (more on this below).



Next, we ask the question: \emph{How can we best arrange the states, operators and measurements, so that the final quantum state maximises the fitness function?} This is a search problem; to perform this search, we use a genetic algorithm, which is a powerful and flexible global search metaheuristic based on evolution by natural selection.

To use a genetic algorithm for our problem, we must first encode each potential experiment into a \textit{genome}. Our genome is a vector containing a mixture of integers and real numbers. For the most part, the integers in the genome encode \textit{which} quantum optics elements will make up the experiment and the \textit{arrangement} of these elements. The real numbers then encode the \textit{parameters} of the different elements, such as the transmission probability of a beam splitter or the magnitude and phase of a coherent state. Given any (valid) genome, AdaQuantum then simulates the corresponding quantum optics experiment, determines what quantum state will be outputted, and calculates the fitness value by applying the chosen fitness function to the output state. See Fig.~\ref{example_schematics} for examples of different experiments, each of which corresponds to a different genome.

\begin{figure}[]
	\centering
	\includegraphics[width=1\linewidth]{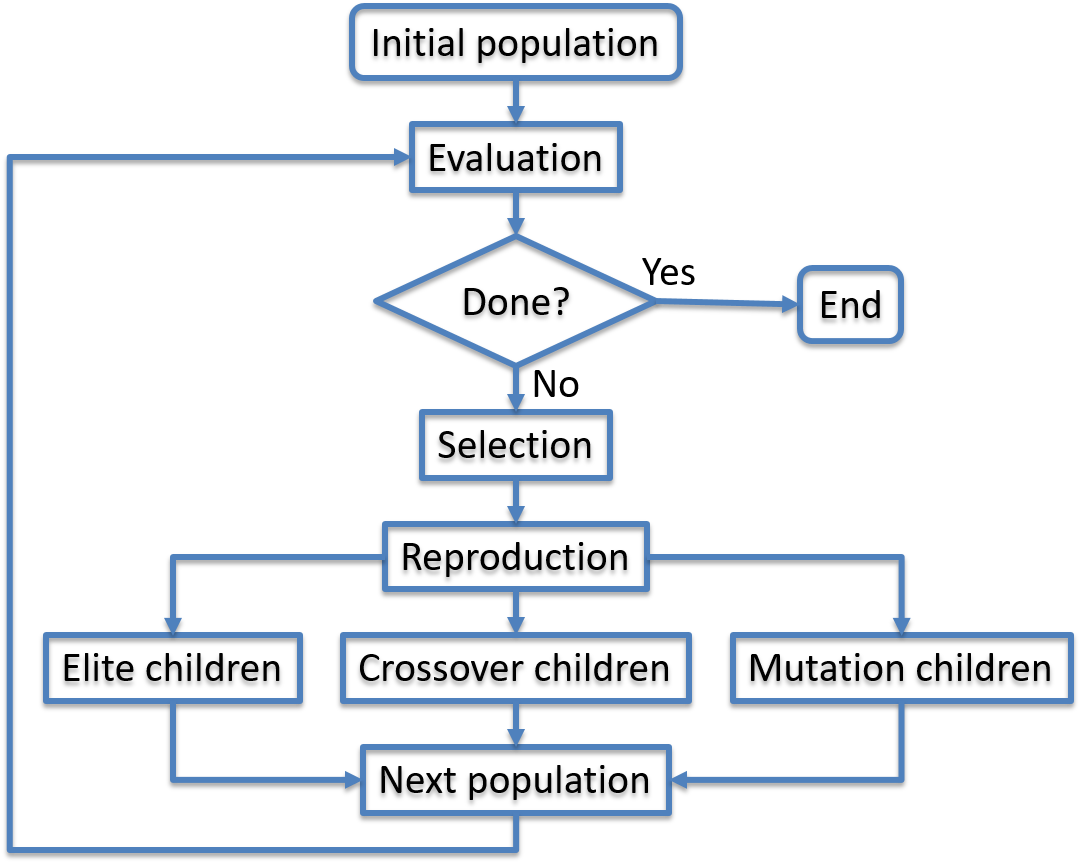}
	\caption{Flowchart of a genetic algorithm -- the various steps are described in the main text.}
	\label{fig:geneticalgorithmflowchart}
\end{figure}

The task we then set the genetic algorithm is to search for a genome that maximises the fitness value. Genetic algorithms start by creating a collection of genomes, which together are known as the \emph{population}. Next, the fitness function for each genome in the population is evaluated. The ``fittest'' genomes -- i.e. the genomes with the largest fitness value -- are then selected, and a new population of genomes is generated by mixing some of the genomes together (\emph{crossover}) and by modifying (\emph{mutating}) others. This next population should, in principle, be comprised of genomes that are ``fitter''. This process repeats through a number of \emph{generations}, until it is unlikely that any more generations will result in improvements. At this stage, if the algorithm has been designed appropriately, the fittest genomes will encode optimised solutions. A flow chart of a genetic algorithm is given in Fig.~\ref{fig:geneticalgorithmflowchart}, and a more detailed description of how genetic algorithms work can be found in Appendix.~\ref{APX:ga_details}.

\subsection{Using a 3-stage algorithm}
\label{sec:three_stage}

The slowest part of running the genetic algorithm is the simulation of each quantum optics experiment, and this must be done a large number of times in order to complete a thorough search of the search space (here the search space is the space of all possible quantum optics experiments that can be constructed with the selected toolbox). The dominant parameter that determines the speed of the simulation is the \textit{truncation} of the Hilbert space: we have to truncate each state $\ket{\psi}$ (and correspondingly the operators and measurements), such that $\ket{\psi} = \sum_{i=0}^{\infty} c_i \ket{i}$ becomes $\ket{\psi} \approx \sum_{i=0}^{T} c_i \ket{i}$, where $T$ is the truncation. The larger the value of $T$ the more accurate the simulation will be, but the time taken for each simulation increases (approximately) exponentially as we increase $T$.

However, while the ``best'' quantum states we found required a large value of $T$ in order for the simulation to be accurate (e.g. see Fig.~\ref{fig:number_dist}), we also found that for a significant proportion of the possible experiments an approximate simulation, using a small value of $T$, can still provide valuable information to guide the search. To exploit this fact, and overcome the challenge of requiring a large $T$ for accuracy but a small $T$ for speed, we settled upon the following 3-stage algorithm:\\

\noindent Stage 1:	A large number (around $10^7$ for our main algorithm runs) of random genomes are created and evaluated using simulations with a small $T$, e.g. $T=30$. A collection of genomes with the best fitness values are selected for the next stage. While this first stage is only approximate due to the small value of $T$, it still rules out many ineffective experiments and gives the genetic algorithm in the next stage a much stronger starting population. \\

\noindent Stage 2:	We run Matlab's inbuilt genetic algorithm \cite{matlabga} with a medium-sized population (around $10^5$ for our main algorithm runs). Here $T$ is larger (e.g. $T=80$) and the genetic algorithm runs for a set number of generations, usually 10.\\


\noindent Stage 3:	In the final stage, the simulation is accurate but slow. In this stage, the fitness function will first simulate the circuit specified by the input genome at a very low truncation (e.g. $T=20$), and then repeat this, increasing $T$ on each iteration (in steps of 10), until both the mean number of photons in the final state and the fitness value converge (indicating that increasing $T$ does not change our results), or until the maximum value of $T$ is reached (where the maximum value of $T$ is specified by the user, labeled ``Max. Truncation'' in Fig.~\ref{fig:UI - toolbox}). This ensures the results are reliable and accurate, while running much faster than if we had chosen the maximum value of $T$ for each genome. Here the population is smaller (around $10^4$ for our main algorithm runs). By the end of this stage (which usually runs for around 40 generations), the algorithm should converge to a genome corresponding to a quantum optics experiment that produces the desired quantum state. \\

We compared our 3-stage genetic algorithm to Matlab's built-in standard genetic algorithm, pattern search, swarm, and simulated annealing algorithms \cite{matlabga}, and our algorithm performed significantly better than all. Our tests also showed that removing any one of the 3 stages described above gave sub-optimal results. In Appendix \ref{APX:running_for_results} we overview the genetic algorithm settings (such as the selection, mutation and crossover functions), and give details of the hyperparameters we use.

We took a number of steps to speed up our algorithm \footnote{Among other techniques, we used Matlab's sparse matrix feature, and used analytical calculations to speed up a number of steps.}. Most importantly, we utilised the algorithm outlined in \cite{al2011computing}, and available as a Matlab function at \cite{expmvMatlab}, which implements our operations by calculating $\exp(tA)b$ without needing to explicitly form $\exp(tA)$, where $t$ is a scalar, $A$ is a matrix and $b$ is a vector. This produces a dramatic speed up, and allows us to simulate experiments with a value of $T$ as large as $170$ photons (in two modes) in the order of seconds. This is a significantly larger value of $T$ than comparable algorithms in the literature \cite{knott2016search,killoran2018strawberry}. As we will see next, this allows us to generate, among other things, quantum states comprised of a superposition of the vacuum with $80$ photons.

\subsection{Designing AdaQuantum for flexibility}
\label{sec:flexibility}

A key design focus when constructing AdaQuantum was flexibility, so that researchers with diverse requirements can use it to achieve their goals. The first flexible aspect is the toolbox selection, which is done using the user interface in Fig.~\ref{fig:UI - toolbox}. Once the user has selected the states, operators, and measurements, AdaQuantum automatically designs a genome that reflects this choice. Therefore, different runs of the algorithm with different toolbox selections will have different genome structures -- both the genome length and the positions of the integers and real numbers will change. The number of modes in the experiment can also be modified, as can the number of operations prior to measurement (see Fig.~\ref{fig:stateengineeringscheme}), and again AdaQuantum adapts to this. For example, when we select 4-modes the algorithm automatically knows that the input state can contain at most two 2-mode states, and that we will require three heralding measurements in order to produce a single mode output state.

In addition, when the user chooses a loss-free experiment with either number resolving or homodyne detectors, then the output state will be a pure state, whereas if there is loss (on either the final state or the measurements), or if the measurements have uncertain outcomes (i.e. when using the on/off or multiplex detector), then the output will be mixed. These features allow a huge range of experiments to be searched over, with applications for both experimentalists looking for lab-ready experiments, and theorists looking to push the limits of what is possible in hypothetical experiments.

Another key part of the flexibility is that AdaQuantum has been designed so that elements can be easily added, removed or modified from the toolbox, and similarly for the fitness functions. AdaQuantum is available on GitHub \cite{AdaQuantum_GitHub}, which includes a User Guide that explains how to modify the toolbox and fitness functions. Modifying the toolbox will allow experimental elements that we haven't yet considered to be included, increasing the applicability of the algorithm. Perhaps more importantly, adding more fitness functions will allow AdaQuantum to be used to find quantum states for a wide range of applications. As long as a fitness function can be programmed in a form that takes as input a density matrix, and outputs a real number, then it can be incorporated into the algorithm.

Once a fitness function or toolbox element is added/modified, AdaQuantum automatically updates the user interface (Fig.~\ref{fig:UI - toolbox}), and again the genome changes accordingly. Therefore, with simple modifications AdaQuantum can, for example, search for non-Gaussian states \cite{straka2018quantum,takagi2018convex}, states for multi-parameter estimation \cite{humphreys2013quantum,vidrighin2014joint,szczykulska2016multi,Proctor2017}, or states with certain non-classical properties \cite{lee2016sub,zhu2016quasiprobability,sperling2017identification}. In this paper we focus on fitness functions for quantum-metrology applications (see next section), and in a spin-off project we have already designed fitness functions -- and then used AdaQuantum to design experiments -- to produce states with a high fidelity to a range of target states, such as cat states \cite{o2018hybrid}.

\section{Results}
\label{sec:results}

In this section, we present the results of running the algorithm with several different fitness functions and toolboxes, which are all relevant to the field of quantum metrology. In Appendix \ref{APX:running_for_results} we give a detailed discussion of the numerous steps and choices involved in running AdaQuantum to obtain these results.

\begin{figure}
\centering
\includegraphics[trim=0.4cm 6cm 1cm 7cm, clip=true, totalheight=0.3\textheight]{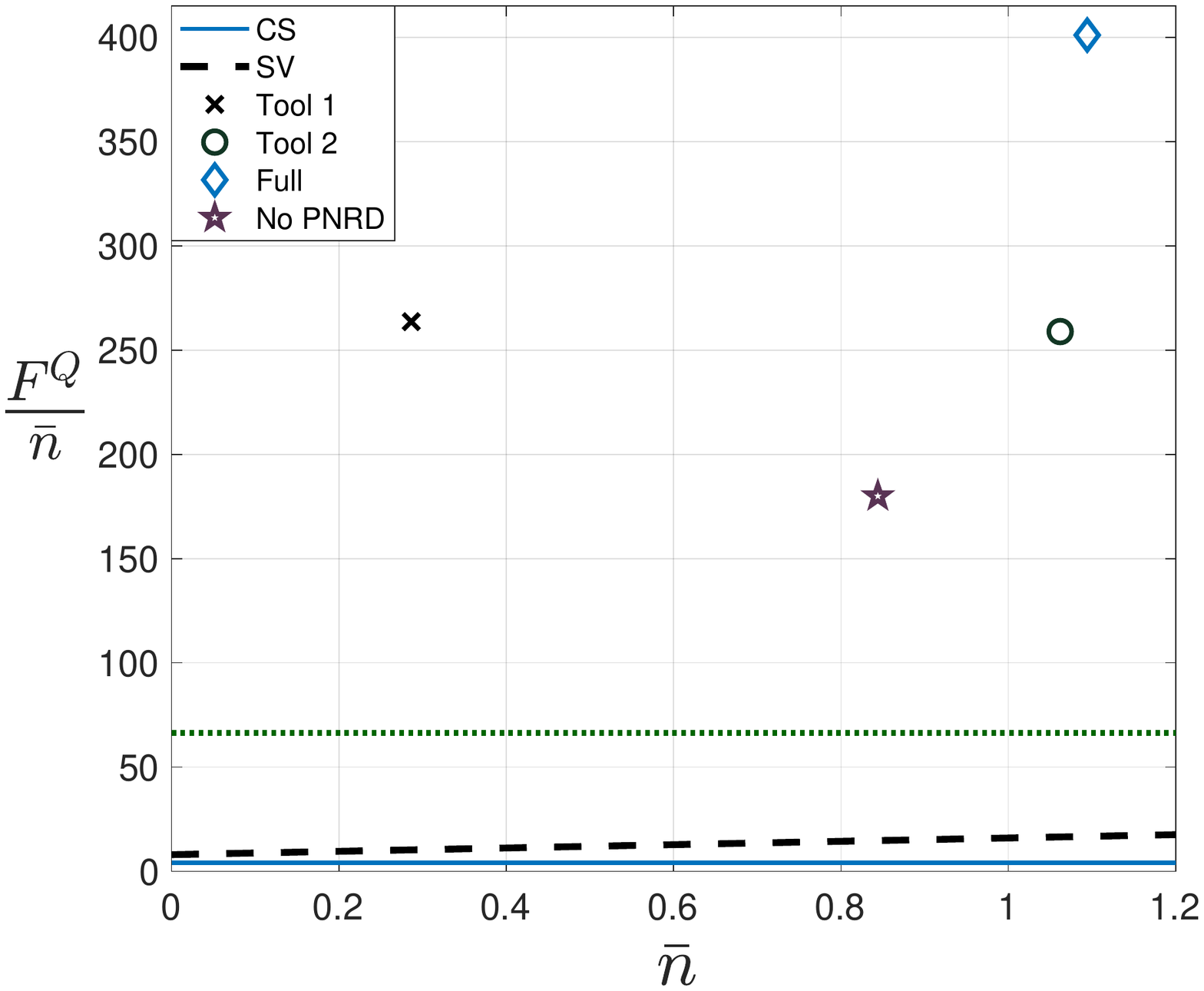}\\
		\caption{Noiseless results: The QFI (for phase estimation), scaled by the mean photon number, for each of the states produced by our algorithm, compared to the squeezed vacuum state (SV) and coherent state (CS). For comparison, the value of ${F^Q}/{\bar{n}}$ for the coherent state is $4$. See Table.~\ref{results_noloss_table} for details of all the states. The toolboxes \emph{Full}, \emph{Tool 1}, \emph{Tool 2}, and \emph{No PNRD} are described in the main text. We see large improvements over the CS and SV, even for the experimentally-viable toolboxes \emph{Tool 1}, \emph{Tool 2}, and \emph{No PNRD}. We also compare our results to the squeezed vacuum state with the largest value of squeezing so far produced in experiments, which, to the best of our knowledge, is the $15$dB squeezed state in Ref.~\cite{vahlbruch2016detection}. For this state, ${F^Q}/{\bar{n}} = 66.4$, which is illustrated as the horizontal dotted green line.}
		\label{fig:Fovernbar_vs_nbar}
\end{figure}

\subsection{Fitness function 1: Pure state QFI}
\label{sec:QFI}

For our first fitness function we consider only noise-free experiments, and we consider the quantum Fisher information (QFI) of the (pure) output state $\ket{\psi}$ for measuring an unknown phase $\phi$. This is given by \citep{braunstein1994statistical}
\begin{equation}
\label{equ:QFI}
F^Q = 4(\braket{\psi_{\phi}'|\psi_{\phi}'} - |\braket{\psi_{\phi}'|\psi_{\phi}}|^2).
\end{equation}
Here $\ket{\psi_{\phi}} = \exp(i\hat{n}\phi)\ket{\psi}$ is our state after the phase shift $\phi$ is applied, and $|\psi_{\phi}'\rangle \equiv {\partial \over \partial \phi} \ket{\psi_{\phi}}$. The QFI is useful because it can be used to tell us the precision with which our state $\ket{\psi}$ can measure the phase shift $\phi$. This is found using the quantum Cram\'er-Rao bound (CRB) \citep{braunstein1994statistical}:
\begin{eqnarray}
\delta \phi \ge \frac{1}{\sqrt{\mu F^Q}}, \label{eq:CRB}
\end{eqnarray}
where $\delta \phi$ is said precision, and $\mu$ is the number of repetitions of the experiment. 

Note that $|\psi_{\phi}'\rangle = i\hat{n}|\psi_{\phi}\rangle$. We can therefore write Eq.~(\ref{equ:QFI}) independently of $\phi$, for example
\begin{align}
| \braket{\psi_{\phi}' | \psi_{\phi}} |^2 &= | \braket{\psi_{\phi} |\hat{n}| \psi_{\phi}} |^2 \\
&= | \braket{\psi |\hat{n}| \psi} |^2
\end{align}
In this way, we can calculate the QFI directly from the output state of our circuit, $\ket{\psi}$, without needing to perform any differentiation, which improves the performance of our algorithm.

The actual fitness function we use here is the QFI scaled by $\bar{n}$ (the mean photon number of the output state), ${F^Q}/{\bar{n}}$. If we just used the un-scaled QFI, then the algorithm would just try to make bigger and bigger states, as even a classical state has an unbounded QFI in the large photon-number limit. Such an optimisation would soon produce states that require Hilbert-space truncations that are both too large to simulate, and require unrealistic experimental equipment. Using ${F^Q}/{\bar{n}}$ overcomes this problem. (Note that it would also be possible to include a restraint in the fitness function that penalizes the deviation of $\bar{n}$ from a given number.)

We also use ${F^Q}/{\bar{n}}$ to compare our results with alternatives in the literature. While a large QFI does amount to a high precision, it does not give a fair comparison when different states have different intensities (where we define intensity as the mean number of photons in the state). To compare different states the key quantity of interest is the precision that a given state can estimate the phase shift ($\delta\phi$), but in an attempt to keep the comparison fair we choose to fix the total intensity that each state is allowed to use in the estimation procedure. We label the total intensity (the `resources') $R$, which is given by $R=\bar{n}\mu$, where $\mu$ is the number of times the state is sent through the phase shift. By taking $R$ as a constant, the CRB in Eq.~(\ref{eq:CRB}) can be written as
\begin{equation}
\label{eq:gary_gary}
\delta\phi \geq \sqrt{\frac{\bar{n}}{R F^Q}} \propto  \sqrt{\frac{\bar{n}}{F^Q}}.
\end{equation}
We see that by fixing $R$, we can use ${F^Q}/{\bar{n}}$ to compare the performance of different states.

To judge the success of our results, we will compare the states we produce to a coherent state, which is the optimal \emph{classical} state, and to a squeezed vacuum state, which is the optimal Gaussian state when there is no noise \cite{Monras2006, de2015two}. Fig.~\ref{fig:Fovernbar_vs_nbar} shows the comparison of these states with our results for the noiseless experiments found by AdaQuantum. Here we chose to display the squeezed vacuum and coherent states by plotting $\bar{n}$ against ${F^Q}/{\bar{n}}$. But this is not the only way to compare these states with our results. As an alternative, we can compare the states found by AdaQuantum against a squeezed vacuum state with the largest value of squeezing so far produced in experiments, which, to the best of our knowledge, is the $15$dB squeezed state in Ref.~\cite{vahlbruch2016detection}. For this state, ${F^Q}/{\bar{n}} = 66.4$ (calculated using \cite{sahota2015quantum}, with $|\zeta|=1.73$), which we include in Fig.~\ref{fig:Fovernbar_vs_nbar} as the horizontal dotted green line.

\begin{figure*}[t]
	\centering
	\subfigure
 {
	\includegraphics[width=0.4\linewidth]{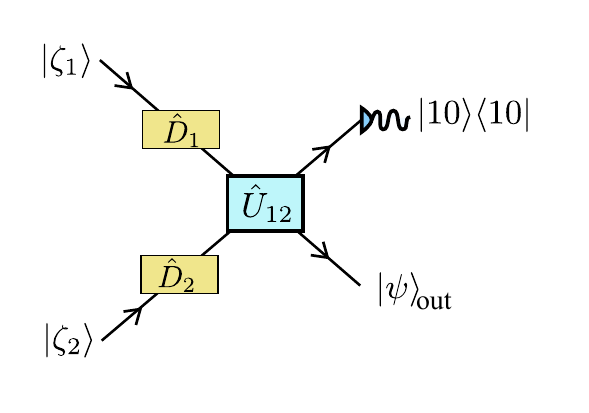}
	\includegraphics[width=0.4\linewidth]{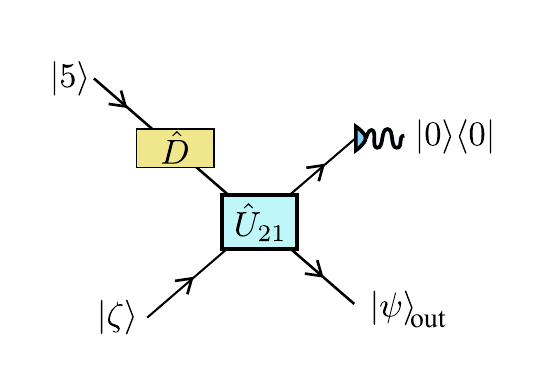}
	}
	\subfigure
	{
	\includegraphics[width=0.4\linewidth]{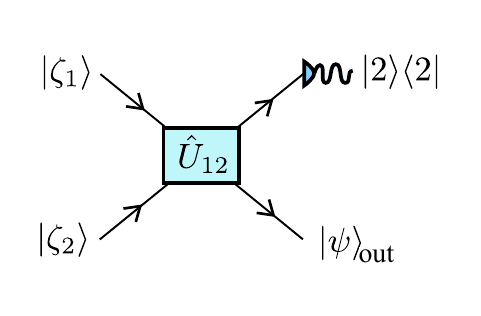}
	\includegraphics[width=0.4\linewidth]{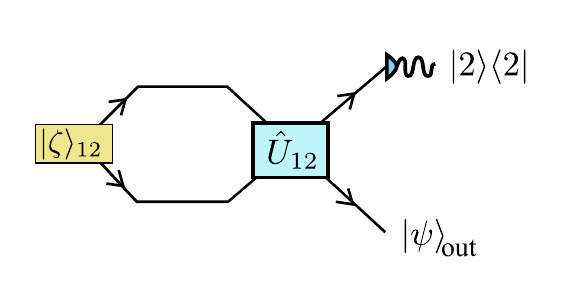}
	}
	\caption{A selection of the experiments designed by AdaQuantum. Top left and top right: These designs produce states that optimise the QFI, as introduced in Section \ref{sec:QFI}. The performance of these states is shown in Fig.~\ref{fig:Fovernbar_vs_nbar}, and the details of the parameters are given in Table.~\ref{results_noloss_table}. The top left design uses the toolbox \emph{Tool 1}, whereas top right uses toolbox \emph{No PNRD}. Bottom left: A design to produce a state that optimises the QFI, after loss is applied to the measurement and final state, as introduced in Section \ref{sec:QFI_loss}. The performance of this state is shown in Fig.~\ref{fig:stateeng_loss}, where each loss rate requires different parameters. Bottom right: A design to produce a state that optimises the Bayesian mean square error (BMSE) with $1$ repetition ($\mu = 1$), as introduced in Section \ref{sec:BMSE}. The performance of this state is shown in Fig.~\ref{bmse_ada_results}, and the details of the parameters are given in Table.~\ref{toolbox_summary}.}
	\label{example_schematics}
\end{figure*}

\begin{table*} [t]
{\renewcommand{\arraystretch}{1.5} 
\begin{tabular}{lllllll}
Toolbox & $\psi_{in}$ & $\mathcal{O}_1$ & $\mathcal{O}_2$ & $\mathcal{O}_3$ & POVM & $p(\%)$ \\
\hline
Tool 1 & $\ket{ \zeta_1 = 1.39 e^{i2.50} , \zeta_2 = 0.34 e^{i5.64} }$	& $\hat{D}_2(\alpha = 2.49 e^{i5.92})$ & $\hat{D}_1(\alpha = 1.66 e^{i6.11})$ & $\hat{U}_{12}(T = 0.30)$ & $\ket{n = 10}\bra{n = 10}$ & 1.19\\ 
Tool 2 & $\ket{\zeta_1 = 1.40 e^{i2.35} , \ \zeta_2 = 0.31 e^{i5.44} }$ 	& $\hat{D}_2(\alpha = 1.97 e^{i2.72})$ & $\hat{D}_1(\alpha = 2.34 e^{i5.98})$ & $\hat{U}_{21}(T=0.18)$ & $\ket{n = 6}\bra{n = 6}$ & 1.72\\ 		
Full & $\ket{ 0 ,\ \zeta = 1.32 e^{i0.06} }$ 	& $\hat{S}_{12}(\zeta = 0.88 e^{i4.73})$ & $\hat{D}_1(\alpha = 3.20 e^{i6.28})$ & $\hat{S}_{2}(\zeta = 0.19 e^{i6.25})$ & $\ket{n = 10}\bra{n = 10}$ & 3.05\\  
No PNRD	& $\ket{ n = 5 ,\ \zeta = 1.40 e^{i6.09} }$ 	& $\hat{D}_1(\alpha = 2.30 e^{i4.62})$ & $\hat{U}_{21}(T=0.66)$ & --- & $\ket{n=0}\bra{n=0}$ & 8.60\\ 
	\end{tabular}}
\caption{Toolboxes used and circuits produced for the results presented in Fig.~\ref{fig:Fovernbar_vs_nbar} (schematics of a selection of the experiments designed by AdaQuantum are given in Fig.~\ref{example_schematics}). The contents of each of the toolboxes are described in the text, and $p(\%)$ refers to the heralding success probability as a percentage. To clarify the notation used here, the state found with toolbox \emph{Tool 1}, for example, is engineered by first creating a pair of squeezed vacuum states (with the parameters given in the table). Next, displacement operators act on both modes. A beam splitter is then applied, which entangles the two modes, and finally a $10$ photon heralding measurement is performed. The probability of this heralding measurement being successful is $1.19\%$. The algorithm's flexibility means that we could easily re-run it to search for states with a higher heralding probability, if this was desired.}
\label{results_noloss_table}
\end{table*}

In Fig.~\ref{fig:Fovernbar_vs_nbar} a different experimental scheme has been found for each of the four different toolboxes we have tested. The most expansive toolbox we studied is named `Full' in the figure. This includes all of the toolbox elements listed in Table~\ref{fig:toolbox} (and described in Section \ref{sec:experimentalquantumoptics}). Note that, for this toolbox, squeezing operators are allowed on arbitrary states. In this toolbox, the strength of the squeezing $|\zeta|$ is limited to $1.4$ in both the squeezing operators and squeezed states; $|\alpha|$ is limited to $5$ in both initial coherent states and displacement operators; and number states are limited to $n = 5$. The limit to the number of photons resolved by PNRDs is $n= 10$. The best experiment devised by the algorithm, detailed in Table.~\ref{results_noloss_table}, utilises the squeezing operators and heralds on the detection of $10$ photons. The resulting state has a ${F^Q}/{\bar{n}}$ approximately $20$ times higher than the corresponding squeezed vacuum state, and $100$ times higher than the coherent state (and note that the squeezed vacuum state has a larger QFI than both the NOON state and the Holland and Burnett state \cite{sahota2015quantum,lee2002quantum,holland1993interferometric}). This state even beats the largest squeezed vaccum so far produced in experiments (to the best of our knowledge) \cite{vahlbruch2016detection} by a factor of 6 (though if we had allowed this much squeezing in our runs of AdaQuantum, as opposed to limiting the squeezing to $|\zeta|=1.4$, then our results would likely be improved significantly).

However, acting with squeezing operations on arbitrary states is extremely challenging at present, so we will move on to toolboxes `Tool 1' and `Tool 2'. These are both the same as `Full' but with the squeezing operations removed. They are very similar to each other, the only difference being that `Tool 2' has the photon detection limited at $n = 10$, as in `Full', and `Tool 1' has the photon detection limited to $n = 6$. Running the algorithm with these toolboxes produced states with ${F^Q}/{\bar{n}}$ still far higher than the squeezed vacuum and coherent state. Finally, we remove the ideal number measurements to form the toolbox `No PNRD', which instead includes bucket detectors and multiplex detection consisting of $16$ bucket detectors, for up to $6$ photon heralding. The resulting experiment found by AdaQuantum uses a heralding measurement on the vacuum $\ket{0}\bra{0}$ and produces a state with ${F^Q}/{\bar{n}}$ still significantly higher than the squeezed vacuum. The states found by AdaQuantum for all these toolboxes are detailed in Table \ref{results_noloss_table}, and schematics of a selection of the experiments designed by AdaQuantum are given in Fig.~\ref{example_schematics}.

\begin{figure}
\centering
\includegraphics[trim=2.4cm 8cm 1.5cm 8cm, clip=true, totalheight=0.33\textheight]{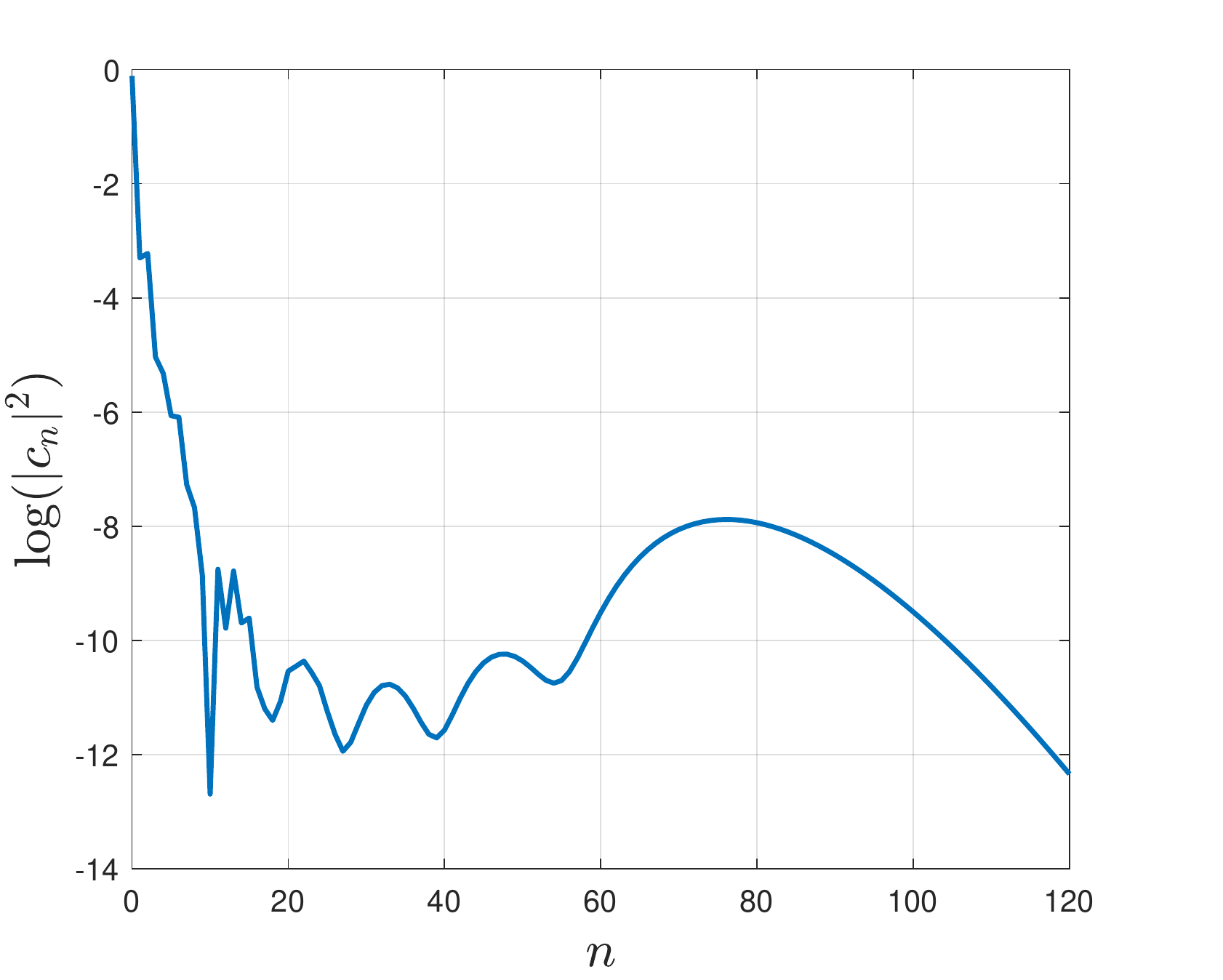}
		\caption{Writing the state generated by `Tool 2' as $\ket{\psi} = \sum_{n=0}^{\infty} c_n \ket{n}$, we plot the log of $|c_n|^2$ against $n$. We see that this state is a superposition of the vacuum with a large number of photons, and hence can be seen as a Schr\"odinger-cat-like state. To see how this state is made, see Table.~\ref{results_noloss_table}.}
		\label{fig:number_dist}
\end{figure}

Why do these states give such large improvements over the alternatives? This can be revealed by looking at the number-distribution of the states. Here we focus on the state labeled `Tool 2', and leave such an analysis of the other states for future work. By writing the state as
\begin{equation}
\ket{\psi} = \sum_{n=0}^{\infty} c_n \ket{n},
\end{equation}
in Fig.~\ref{fig:number_dist} we plot the log of $|c_n|^2$ against $n$ \footnote{To rule out the possibility that the large QFI could be due to numerical inaccuracies, we double-checked the calculation leading to Fig.~\ref{fig:number_dist} analytically. This can be done by writing the displaced-squeezed states in the number basis in terms of creation and annihilation operators, propagating these creation and annihilation operators through a beam splitter, before finally acting on the LHS with $\bra{6} \otimes \mathbb{I}$ and normalising the resulting single-mode state.}. The figure reveals that this state has a large contribution of the vacuum, and a small but significant contribution of a huge number of photons, around $60$-$100$. Such a state can be seen as a Schr\"odinger-cat-like state: the human eye can directly detect less than $100$ photons \cite{rieke1998single}, so, arguably, this state can be seen as a superposition of a macroscopic state with the vacuum. More importantly for metrology, such a state has a large QFI because it has a large variance (with respect to the encoding Hamiltonian), whilst retaining a small mean number of photons. This state can therefore be seen as a realistic version of the so-called ON state, which has already attracted much attention in metrology \cite{rivas2012sub,knott2016local}. This state is also useful for quantum computing with continuous variables \cite{sabapathy2018states}.

\subsection{Fitness function 2: Mixed-state QFI}
\label{sec:QFI_loss}

For our next fitness function, we study the effect of the two most dominant noise sources in the quantum optics experiments we are considering: photon loss at the output and imperfect heralding measurements. After this noise is applied the resulting state will be a mixed state $\rho$. We then apply the mixed-state QFI to this state (again scaled by the mean photon number in the state, $\bar{n}$). To calculate the mixed-state QFI, we first need to apply the phase shift to $\rho$, giving $\rho_{\phi}$, and then find the eigenvalues and eigenvectors of this state by writing it as $\rho_{\phi} = \sum_m q_m \ket{\psi_m}\bra{\psi_m}$. We then use the form of the mixed-state QFI developed in \cite{zhang2013quantum}: 
\begin{equation}
\label{eq:QFI for state eng}
F^Q = \sum_i q_i F^Q_i - \sum_{i \neq j} \frac{8 q_i q_j}{q_i + q_j} |\bra{\psi_i'}\psi_j\rangle|^2 
\end{equation}
where $F^Q_i$ is the pure state QFI of eigenstate $\ket{\psi_i}$. We use a similar technique as with the pure-state QFI to write the mixed-state QFI so that it does not involve differentiation. It should be noted that we deal with pure states, in the form of vectors, until the measurement stage, when we switch to the density matrix formalism, as this allows us to use the matrix exponential technique discussed above and in \cite{al2011computing}.

\begin{figure}[t]
\centering
\includegraphics[trim=3cm 6cm 3cm 6.5cm, clip=true, totalheight=0.4\textheight]{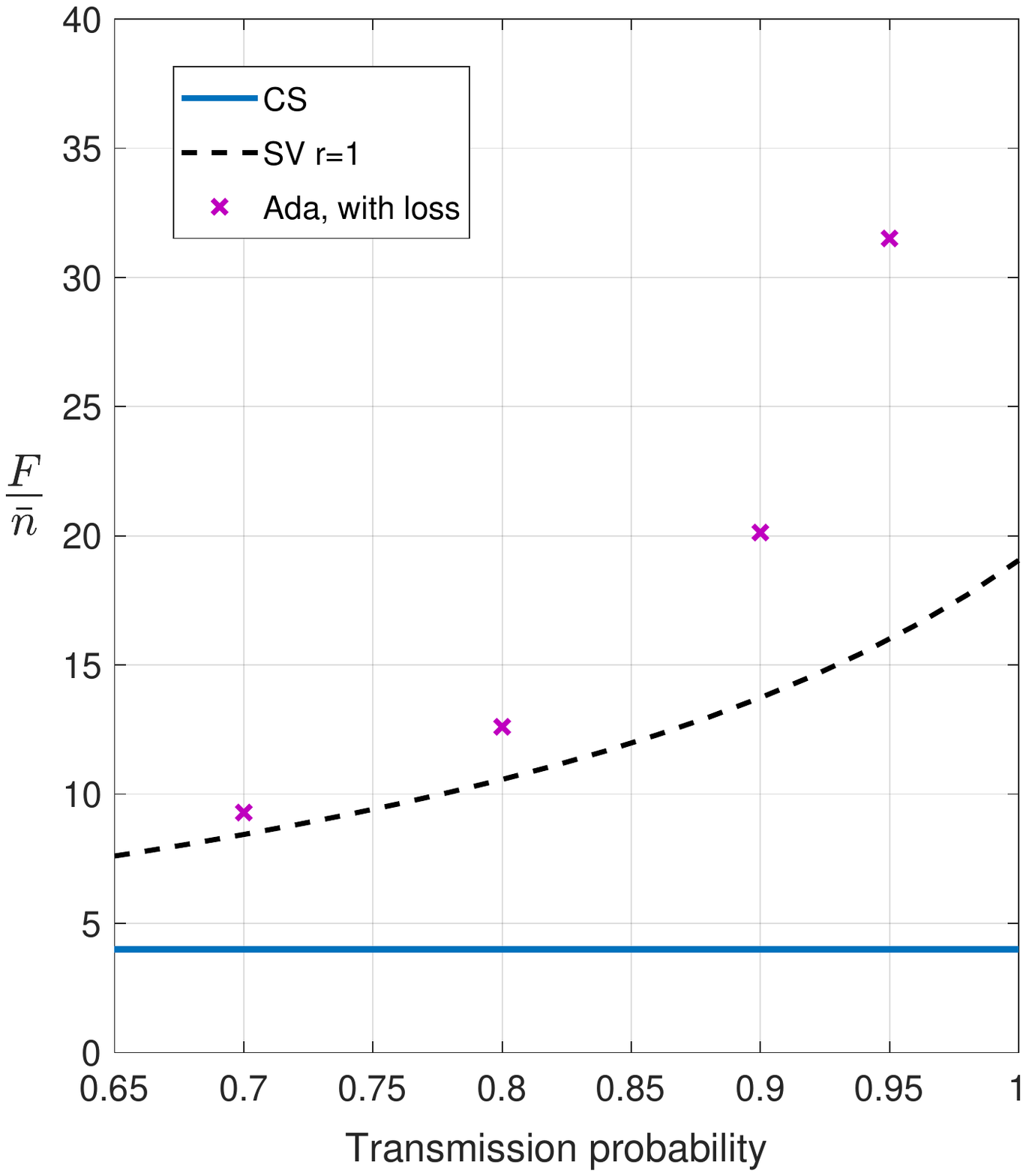}
		\caption{We plot ${F^Q}/{\bar{n}}$ of the states found by AdaQuantum against the transmission probability ($1 -$ loss rate) of both the loss on the output state, and the loss in the heralding measurements. All states found by AdaQuantum take the form $\ket{\psi} = \mathcal{N} \bra{2} \hat{U}_{12} |\zeta_1,\zeta_2\rangle$, where $\mathcal{N}$ is the normalisation, i.e. these states are formed by sending two squeezed vacuum states through a beam splitter, followed by $2$-photon heralding. Note that, despite the fact that the states found by AdaQuantum have the same form for all loss rates, we optimised AdaQuantum for each loss rate separately. I.e. we first fixed the loss rate, then ran AdaQuantum to find states to maximise the QFI for that rate.}
		\label{fig:stateeng_loss}
\end{figure}

In the noiseless case, we saw that AdaQuantum found quantum states with dramatic improvements over the squeezed vacuum and coherent state. Noise will inevitably diminish such improvements, but up to now it has been an open question whether it was possible to improve over the squeezed vacuum and coherent state \emph{at all} when realistic experimental equipment and noise are factored in. Note that the squeezed vacuum and coherent state do not require heralding, so they escape the effects of imperfect heralding.

As discussed above, imperfect heralding can be modeled by applying loss to the state prior to heralding. For simplicity, we fix the loss to be the same for both the heralding measurements and on the output state. Unlike the noiseless case, here we are specifically focused on experiments that can be realistically performed without specialised equipment. Therefore, the toolbox we use here does not use squeezing operations and has the following limits to the parameters: $n = 4$ for Fock states; $|\alpha| = 5$ for coherent states and displacement operator; $|\zeta| = 1$ for squeezed states; and photon number resolving up to $n = 6$.

The results produced for varying values of loss are shown in Fig.~\ref{fig:stateeng_loss}. The value of ${F^Q}/{\bar{n}}$ for each of these results is higher than the squeezed vacuum at the same loss rate -- this opens up the possibility for experiments to create these exotic non-Gaussian states, and use them to beat the squeezed vacuum state in phase estimation. The heralding probabilities to produce the desired output states are quite reasonable (around $10$-$20 \%$) and so the experiments proposed here can be carried out experimentally in reasonable time. Note that this experiment requires a 2-photon heralding measurement (which can be performed with multiplexed bucket-detectors) and single mode squeezed states, which can all be achieved with current or near-future technology (for example see \cite{humphreys2015tomography,afek2010high,matthews2016towards,achilles2004photon,xiang2011entanglement}).

\subsection{Fitness function 3: Beyond the QFI}
\label{sec:BMSE}

The third fitness function that we optimise here is the Bayesian mean square error (BMSE). Despite the importance of the QFI as a method of quantifying the metrological performance of different states, in general its usefulness depends on the possibility of recasting the problem at hand in the language of the local approach to estimation theory \cite{jesus2017, jesus2018, rafal2015, haase2018jul, PARIS2009}. In particular, approaching the CRB in equation (\ref{eq:CRB}) typically requires either having certain prior knowledge and repeating the experiment a large amount of times \cite{jesus2017, berry2012infinite}, or just having a large amount of prior information \cite{haase2018jul, rafal2015}. But in realistic scenarios the number of repetitions of the experiment can be small, and the formalism based on the QFI does not take into account the possibility of having a moderate amount of prior information \cite{jesus2018}. The BMSE factors in both of these, and hence gives a reliable measure of the phase-measuring capability of a given state in a realistic experimental setting.
 
\begin{figure*}
\centering
\includegraphics[trim={0.25cm 0.1cm 1.4cm 0cm},clip,width=9.15cm]{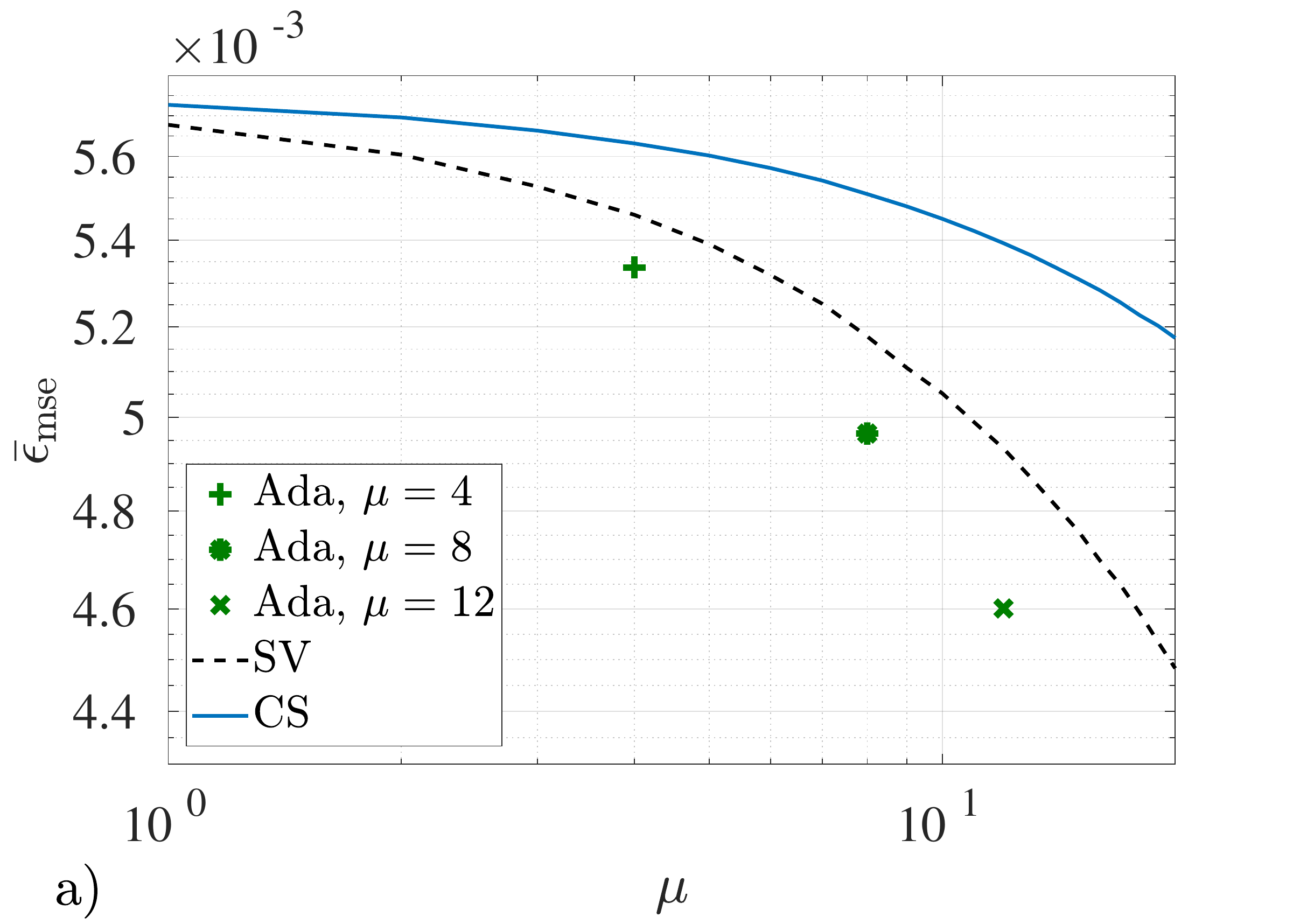}\includegraphics[trim={0.25cm 0.1cm 1.4cm 0cm},clip,width=9.15cm]{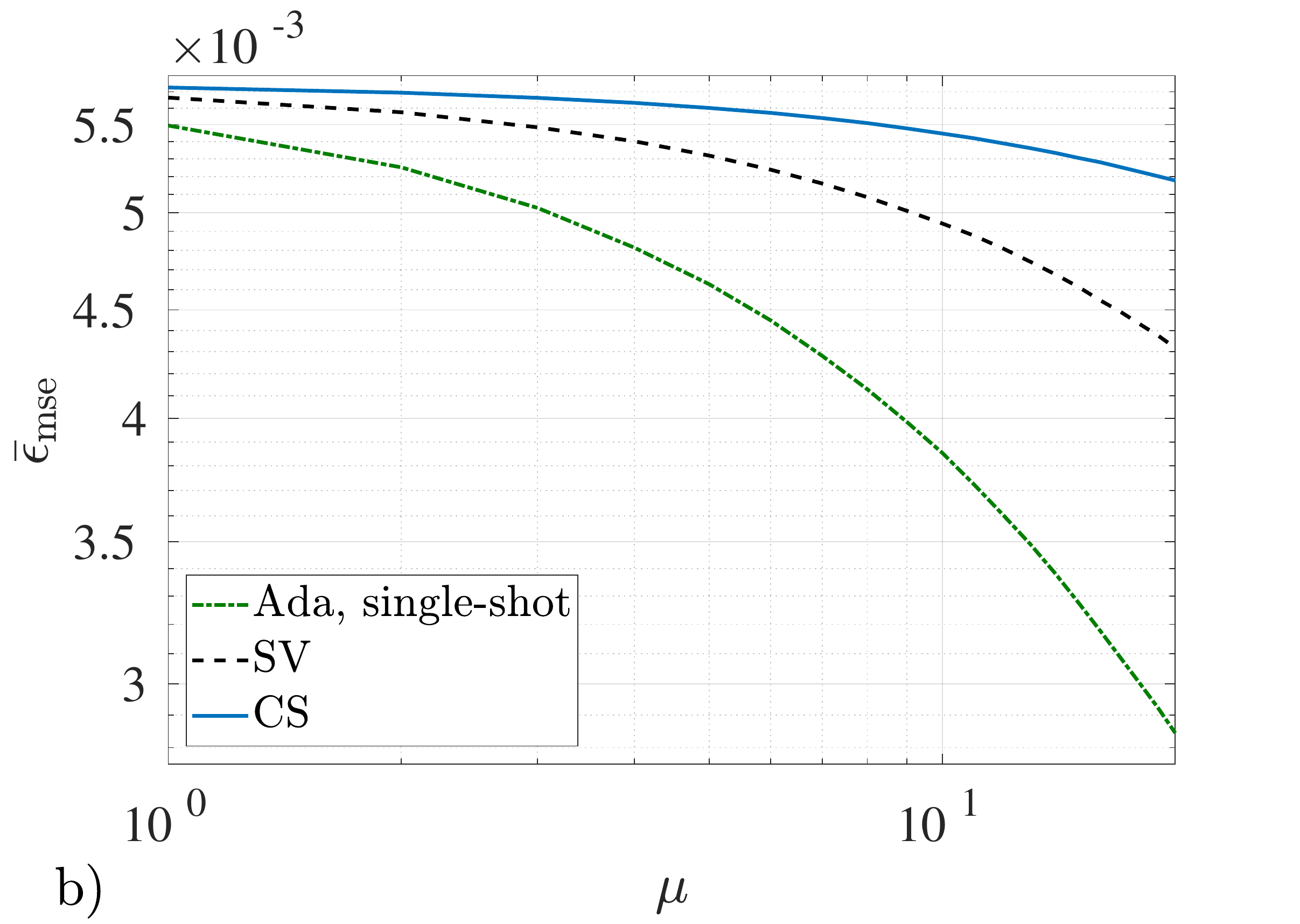}
\caption{Bayesian mean square error as a function of the number of repetitions for a) a coherent state mixed with the vacuum (solid line), two squeezed vacuums (dashed line) and states found by our genetic algorithm, AdaQuantum, for $\mu = 4$ (plus sign), $\mu = 8$ (asterisk) and $\mu = 12$ (cross) repetitions. Here the measurement scheme is based on counting photons after the action of a $50$:$50$ beam splitter. In b) we again consider the coherent state and squeezed state, but now measured by their respective optimal single-shot POVMs. The state found by AdaQuantum (dash-dot line) is then based on the bound in equation (\ref{bayes_bound}), which already takes into account the optimal single-shot POVM for the given state. All the configurations are based on a two-mode interferometer with $1$ photon on average, and where the phase shift can be found in an interval of width $\pi/12$, centred around zero.}
\label{bmse_ada_results}
\end{figure*}

To use the BMSE, unlike when we use the QFI, we need to specify what measurement scheme will be used to extract the information about the phase shift from the probe state. For the results in this paper each run of AdaQuantum produces a \textit{single-mode} state $|\psi \rangle$, which cannot be used on its own to estimate a phase shift because in experiments we can only access the information about relative phase shifts. One approach we can use is to take a pair of such states, $|\psi \rangle \otimes |\psi \rangle$, and then encode a difference of phase shifts as $\ket{\psi_\theta} = \mathrm{exp}[-i(\hat{n}_1-\hat{n}_2)\theta/2]\ket{\psi}\otimes\ket{\psi}$. A measurement can then be performed on both modes, which for simplicity we take to be ideal and given by the POVM elements $\lbrace \ket{M}\bra{M}\rbrace$. Examples of measurements are given below.

The probability of finding the outcome $M$ given the unknown value of the phase is $p(M|\theta) = ||\langle M | \psi_\theta\rangle||^2$, and if the previous preparation and measurement stages are repeated $\mu$ times and the outcomes $\boldsymbol{M} = (M_1, \dots, M_{\mu})$ are recorded, then their probability is $p(\boldsymbol{M}|\theta) = \prod_i p(M_i|\theta)$. Furthermore, imagine that we know in advance that the phase can only be found within an interval of width $\pi/12$ centred around zero, a state of knowledge that can be represented by the prior probability $p(\theta)=12/\pi$, for $\theta \in [-\pi/24,\pi/24]$, and zero otherwise. We can then combine all these pieces of information using Bayes' theorem, which gives us the posterior probability $p(\theta|\boldsymbol{M}) = p(\theta)p(\boldsymbol{M}|\theta)/p(\boldsymbol{M})$, where $p(\boldsymbol{M})$ is the normalisation factor.

At this point we can introduce the BMSE as $\bar{\epsilon}_{\mathrm{mse}} = \int d\boldsymbol{M} p(\boldsymbol{M}) \epsilon(\boldsymbol{M})$, where
\begin{equation}
\epsilon(\boldsymbol{M}) = \left\lbrace\int d\theta p(\theta|\boldsymbol{M}) \theta^2 - \left[\int d\theta p(\theta|\boldsymbol{M}) \theta\right]^2\right\rbrace
\label{variance_mse}
\end{equation}
is the variance of the posterior probability. We note that since the posterior $p(\theta|\boldsymbol{M})$ contains the information about the phase, the BMSE can be understood as an optimal way of quantifying the quality of this information on average for a given POVM. In particular, $\bar{\epsilon}_{\mathrm{mse}}$ is a measure of uncertainty. Further details and the justification of this framework can be found in \cite{jesus2017, jesus2018, rafal2015, jaynes2003}.

The task given to AdaQuantum is, in this case, to find single-mode states $\ket{\psi}$ that minimise $\bar{\epsilon}_{\mathrm{mse}}$ for a given number of repetitions and measurement scheme. Since the calculation of the BMSE is more demanding than finding the QFI (see, e.g., \cite{jesus2017}), we have chosen a narrow prior to simplify the calculation of the integrals in equation (\ref{variance_mse}). However, this does not imply that we could have used the QFI instead, since it was argued in \cite{jesus2018} that the prior width needs to be $0.1$ or smaller if the QFI is to be a suitable figure of merit in a similar scheme, and here the width is $\pi/12 \approx 0.3$. Thus our choice corresponds to the regime of intermediate prior knowledge where the BMSE is useful \cite{jesus2018}. In addition, note that, unlike for the QFI, we do not need to rescale the BMSE by the average photon-number because the BMSE is already a direct measure of the estimation performance. To keep the comparison fair, all states in this section contain an average photon number of 1.

We will optimise the BMSE using two different strategies, both using the same toolbox as for the mixed-state QFI above. First we focus on a specific and practically-motivated POVM: counting photons after the action of a $50$:$50$ beam splitter, and we set the algorithm to optimise the BMSE for $\mu = 4$, $\mu = 8$ and $\mu = 12$ repetitions. This search produces a state that takes the form $\ket{\psi} = \mathcal{N}\langle n|\hat{U}_{12}\hat{S}_{12}|0,0\rangle$ for $\mu = 4$ and $\mu=12$, where $\mathcal{N}$ is the normalisation, while for $\mu = 8$ we find $\ket{\psi} = \mathcal{N}\langle n|\hat{P}_1\hat{U}_{12}\hat{S}_{12}|0,0\rangle$, where $P_1$ is a phase shift in the first mode. Table \ref{toolbox_summary} provides the numerical parameters that generate these states. The uncertainty associated with two copies of these probes for each number of trials has been represented in Fig.~\ref{bmse_ada_results}.a (individual points), and we have also included the BMSE of both an interferometer with a squeezed vacuum per port (dashed line) and the configuration that mixes a coherent state and the vacuum with a $50$:$50$ beam splitter (solid line) \footnote{In \cite{jesus2018} it was argued that if the prior probability is centred around zero and the POVM is based on mixing the light beams and counting photons, then for some probes it is necessary to also shift the phase of the second arm after the unknown parameter has been encoded in order to achieve the optimal BMSE. The calculations presented in this section have already taken this into account as prescribed by \cite{jesus2018}.}. This figure shows that the states found by AdaQuantum perform better than these two benchmarks, which constitutes a clear demonstration that AdaQuantum is able to optimise a Bayesian figure of merit beyond the regime where the QFI is useful. 

More concretely, we can quantify this improvement by introducing the quantity
\begin{equation}
I_{\mathrm{r}}=\frac{\bar{\epsilon}_{\mathrm{r}}-\bar{\epsilon}_{\mathrm{ada}}}{\bar{\epsilon}_{\mathrm{r}}},
\label{improvement_factor}
\end{equation}
where $\bar{\epsilon}_{\mathrm{r}}$ is the BMSE of any of the two reference states that we are employing and a positive $I_{\mathrm{r}}$ indicates that there has been an improvement. Its calculation, whose results are summarised in table \ref{bayesian_ada_improvement}, shows an enhancement of between $2\%$ and $7\%$ with respect to the squeezed vacuum, and between $5\%$ and $15\%$ with respect to the coherent state.

\begin{table} [t]
{\renewcommand{\arraystretch}{1.2} 
\begin{tabular}{|c|c|c|c|c|c|}
\hline
\multicolumn{6}{|c|}{AdaQuantum's relative enhancement} \\
\hline
\hline
\multicolumn{4}{|c|}{$50$:$50$ beam splitter \& photon counting } & \multicolumn{2}{c|}{Single-shot POVM} \\
\hline
Ref. & $I_{\mathrm{r}}(\mu=4)$ & $I_{\mathrm{r}}(\mu=8)$ & $I_{\mathrm{r}}(\mu=12)$ & Ref. & $I_{\mathrm{r}}(\mu=1)$ \\
\hline
SV & $0.02$ & $0.04$ & $0.07$ & SV & $0.03$ \\
CS & $0.05$ & $0.10$ & $0.15$ & CS & $0.04$ \\
\hline
\end{tabular}}
\caption{Improvement factor as defined in equation (\ref{improvement_factor}) to quantify the enhancement of the states found by AdaQuantum with respect to two squeezed vacuums (SV) and a coherent state mixed with the vacuum (CS). The details of the experimental configuration are those indicated in Fig.~\ref{bmse_ada_results} and in the main text.}
\label{bayesian_ada_improvement}
\end{table}

\begin{table*} [t]
{\renewcommand{\arraystretch}{1.5} 
\begin{tabular}{lllllll}
Setting & $\psi_{in}$ & $\mathcal{O}_1$ & $\mathcal{O}_2$ & $\mathcal{O}_3$ & POVM \\
\hline
BMSE, $\mu = 8$ & $|0,0\rangle$ & $\hat{S}_{12}(\zeta = 0.89~\mathrm{e}^{i 0.031})$ & $\hat{U}_{12}(T=0.69)$ &  $\mathrm{e}^{i \hat{n}_1 0.32}$ & $| n = 4 \rangle \langle n = 4 |$  \\
BMSE, $\mu = 4, 12$ & $|0,0\rangle$ & $\hat{S}_{12}(\zeta = 0.91~\mathrm{e}^{i 0.040})$ & $\hat{U}_{12}(T=0.66)$ & --- & $| n = 6 \rangle \langle n = 6 |$ \\
BMSE, $\mu = 1$ & $|0,0\rangle$ & $\hat{S}_{12}(\zeta = 0.95~\mathrm{e}^{i 6.1})$ & $\hat{U}_{12}(T=0.72)$ & --- & $| n = 2 \rangle \langle n = 2 |$ \\
\end{tabular}}
\caption{Details of the circuits produced by AdaQuantum using the Bayesian framework. The first two are for a photon counting measurement after the beam splitter and that the last one is for the optimal single-shot POVM.}
\label{toolbox_summary}
\end{table*}

The second strategy is to perform an analytical optimisation over all possible POVMs first, and then set AdaQuantum to find experiments that optimise this strategy. (Note that during AdaQuantum's search, the optimal measurement scheme for estimating the parameters has been selected -- AdaQuantum just optimises over the state-engineering part of the experiment, i.e. the arrangement of elements in Fig.~\ref{fig:stateengineeringscheme}). Following \cite{jesus2018}, first we recall that the single-shot BMSE satisfies \cite{personick1971, helstrom1976, macieszczak2014bayesian}
\begin{equation}
\bar{\epsilon}_{\mathrm{mse}}(\mu=1)\geqslant\int d\theta p(\theta)\theta^2 - \mathrm{Tr}\left(\bar{\rho} S\right),
\label{bayes_bound}
\end{equation}
where $S \rho + \rho S = 2 \bar{\rho}$, $\rho = \int d\theta p(\theta)\ket{\psi_\theta}\bra{\psi_\theta}$ and $\bar{\rho} = \int d\theta p(\theta)\theta\ket{\psi_\theta}\bra{\psi_\theta}$.
This bound can always be saturated when the measurement scheme is given by the projections onto the eigenstates of $S$ \cite{personick1971, helstrom1976, macieszczak2014bayesian}, where $S$ is an operator that only depends on the transformed state and the prior. Therefore, we can set the algorithm to search for states that minimise $\bar{\epsilon}_{\mathrm{mse}}(\mu=1, \ket{M}\bra{M}=\ket{s}\bra{s})$, where $\lbrace \ket{s} \rbrace$ are the eigenvectors of $S$. Here we find another state with the form  $\ket{\psi} = \mathcal{N}\langle n|\hat{U}_{12}\hat{S}_{12}|0,0\rangle$ but with different parameters (see table \ref{toolbox_summary}).

As table \ref{bayesian_ada_improvement} shows, the state found by AdaQuantum for the optimal single-shot measurement is $3\%$ better than two squeezed vacuums measured by their correspondent single-shot POVM, and $4\%$ better than the coherent states. In addition, we note that using this measurement scheme in a sequence of repeated experiments is an appropriate strategy when we cannot or we do not wish to correlate different trials \cite{jesus2018}. The performance of this state for the first $20$ repetitions of the scheme, which has been represented in Fig.~\ref{bmse_ada_results}.b, shows that the state found by AdaQuantum using the optimal single-shot POVM is better than the benchmarks even when the number of repetitions grows. To summarise, we can say that the combination of AdaQuantum and the methodology introduced in \cite{jesus2018} provides a robust method to find practical probe states with a strong performance for those systems that operate in the regime of limited data.

\section{Discussion and conclusions}
\label{sec:discussion}

This paper is not the first to apply techniques from machine learning to the task of designing quantum experiments \cite{knott2016search,krenn2016automated,melnikov2018active,arrazola2018machine,sabapathy2018near} -- we give a full comparison to these alternatives in Appendix \ref{APX:comparison}.

The results in the final section on the BMSE may have important general consequences for quantum metrology. The formal solution to the optimisation problem posed by a general quantum estimation scheme has been known for some time (see \cite{helstrom1976}). However, finding an analytical form of this solution for specific problems is challenging and generally not possible, which explains why we usually rely on bounds such as the CRB. And while in a sense the limits generated by the latter can be seen as fundamental \cite{haase2018jul}, it can be argued that this is only useful when the bound can be safely applied to a problem in practice, which is not always the case (because realistic experiments put limits on the number of experiment repetitions and the prior knowledge available) \cite{jesus2017, berry2012infinite}. Therefore, the fact that our algorithm is able to find useful metrology protocols with more general fitting functions such as the BMSE or its single-shot optimum in equation (\ref{bayes_bound}) allows us to see it as a promising route to design quantum experiments using more reliable and general figures of merit, including, for instance, other cost functions different from the square error, or even multi-parameter systems \cite{Humphreys2013,Proctor2017}.\\

One clear direction for future work is to extend our runs of the algorithm to more than $2$ optical modes. We attempted to do this in this project, but the additional simulation time required for $>2$ optical modes meant that our global search was not effective. For $2$ optical modes, most of the results presented here were generated by running the algorithm for $96$ hours on $16$ cores of the University of Nottingham's High Performance Computing facility. Running for such times allowed us to run the genetic algorithm with very large populations (in the order of thousands or tens of thousands). When moving to $3$ modes, the exponential slow-down in computing time required to simulate this larger Hilbert space meant that such large populations were not possible. To overcome this, one main focus of our future work will be to significantly enhance the global search, by, among other things, exploring a range of metaheuristic search methods, and by improving the way the genome is encoded. We are confident that this will allow effective searches for quantum experiments in \emph{at least} $3$ modes, with the potential of finding quantum states with significant enhancements.

A further direction is to extend the fitness functions optimised by the algorithm. Ongoing work is searching for experiments to produce a range of specific states (such as squeezed cat states \cite{knott2016practical} and GKP states \cite{gottesman2001encoding}). Beyond this, there are a broad range of fitness functions that could easily be incorporated, including searching for non-Gaussian states, highly entangled states, and states with a negative Wigner function (our GitHub page \cite{AdaQuantum_GitHub} includes a user guide that explains how new fitness functions can be added).

Another interesting question for future work is to understand \textit{why} the experiments we presented here perform so well in maximising their respective fitness functions. This might be especially revealing in the states that are robust to noise: why do these specific experimental arrangements found by AdaQuantum create states that still perform well with photon loss? \\

In conclusion, we have introduced a genetic algorithm for designing quantum optics experiments. We demonstrated the flexibility of our algorithm by optimising three different fitness functions, and in all three the algorithm found improvements over the commonly used alternatives. Perhaps most notably, the algorithm found a realistic method of producing a Schr\"odinger-cat-like state comprised of a superposition of the vacuum with a large number of photons (around $80$), which displays substantial phase-measuring improvements over the competing strategies. We emphasise here that our algorithm can in principle be used to design experiments to produce optical states with \emph{any} desired properties, so long as these properties can be quantified and hence optimised over. Our algorithm, together with related work in the literature \cite{knott2016search,krenn2016automated,melnikov2018active,arrazola2018machine,sabapathy2018near}, highlights the power of utilising methods from artificial intelligence and machine learning to design and optimise quantum experiments. \\

\noindent \emph{Acknowledgements:} We acknowledge helpful discussions with Gerardo Adesso, Jacob Dunningham, Ryuji Takagi, Ender {\"O}zcan, Chris Wade and Guillaume Thekkadath. JR acknowledges support from the South East Physics Network (SEPnet). LM was supported by the Bristol Quantum Engineering Centre for Doctoral Training, EPSRC Grant No. EP/L015730/1. JCFM acknowledges support from an EPSRC Early Careers Fellowship (EP/M024385/1) and the EPSRC UK Quantum Technology Hub in Quantum Enhanced Imaging (EP/M01326X/1). PK acknowledges support from the Royal Commission for the Exhibition of 1851.

\subsection*{References}


\bibliographystyle{apsrev}
\bibliography{Rosanna_references_plus}

\begin{widetext}
\appendix

\section{Optical quantum state engineering}
\label{APX:experimentalquantumoptics}

\subsection{Number states and number measurements}
\label{sec:number}

The first input state we include is the number state, or Fock state, $ \ket{n} $. Typically, an experiment would be restricted to number states of low $n$, as generating a state with a large number of photons is technically challenging, so we restrict our simulations to have at most $n = 5$ (though this can easily be modified). As an example, a single photon state can be produced using spontaneous parametric down-conversion \cite{kwiat1995new,lamas2001stimulated,simon2003theory}. In this process, a photon of a particular frequency may be spontaneously converted into two identical photons, each with half of the original frequency. A heralding measurement is then used to detect one photon in this pair, which signals the existence of the other (this can be extended for $n > 1$).

The first measurements we include is measuring the number of photons in a mode -- \emph{photon number resolving detection} (PNRD) -- which is implemented with the POVM element $\ket{n}\bra{n}$. However, it can be difficult to precisely measure the number of photons so we also include two more-easily-implementable measurements. The first of these is a \emph{bucket detector} (aka \emph{on/off}, \emph{threshold}, or \emph{click} detector). This measures whether at least one photon is absorbed by the detector (it `clicks'), but it is unable to determine the exact number of photons. It is represented by the POVM, $\{\ket{0}\bra{0},\mathbb{I} - \ket{0}\bra{0} \}$, where $\mathbb{I}$ is the identity. These are commonly found in laboratories, for example in the form of avalanche photodiodes \cite{Cova96APDs}.

If exact PNRDs are not available, an approximate photon number measurement can be implemented by using several bucket detectors in a \emph{multiplex detector}. This is achieved by separating the mode you want to measure (the target mode) into several modes via spatial- or time-multiplexing. The former can be implemented via a series of beam splitters, as in \cite{xiang2011entanglement}, or by optical fibre splitters, as in \cite{matthews2016towards}. The latter can be implemented using a series of interferometers with paths of varying length, as in \cite{achilles2004photon}. A bucket detection is then performed on each of these separated modes. A perfect PNRD is obtained in the infinite limit of multiplexed bucket detectors, but otherwise there will be a non-zero probability that more than one photon enters the same bucket detector in a given run, resulting in uncertainty in the exact number of photons entering the detector.

We use the following POVM, from \cite{matthews2016towards}, to simulate this multiplex detection, which takes into account all possible coincidence detection patterns. Here, $r$ is the number of bucket detection events and $c$ is the actual number of photons in the target mode. The weights $w_r(c)$ are non-negative and satisfy $ \sum_r  w_r(c) = 1 $. These rely on the number of bucket detectors, $d$, and on $S(c,r)$, which is the Stirling number of the second kind, which counts the number of ways of partitioning $c$ objects into $r$ non-empty subsets. More details on constructing this POVM are found in the Supplemental Material of \cite{matthews2016towards}, which uses methods from \cite{sperling2012true} to compute the weights. The POVM elements $E_r$ are given by
\begin{equation}
\label{eq:multiplex POVM}
E_r = \sum_{c\geq r}^\infty w_r(c) \ket{c}\bra{c}, \quad w_r(c) = \frac{d!S(c,r)}{(d-r)!d^c}.
\end{equation}
In our simulations we use a multiplex detector consisting of either five \cite{xiang2011entanglement} or sixteen \cite{matthews2016towards} bucket detectors, but again this can be easily altered (the choices to use \textit{five} and \textit{sixteen} bucket detectors wasn't due to any fundamental significance to these numbers, but rather because they have been used in experiements \cite{xiang2011entanglement,matthews2016towards}).

\subsection{Coherent states and displacement operators}
\label{sec:coherent}

Coherent states are among the most readily available states as they are produced by a stabilised laser operating well above threshold \cite{barnett2002methods}. To use them in our toolbox, we must construct them in the Fock basis. They can be defined by the action of the displacement operator on the vacuum, $ \ket{\alpha} = \hat{D}(\alpha)\ket{0} $, where
\begin{equation}
\label{eq:disp operator fock basis}
\hat{D}(\alpha) = \exp{ (\alpha \hat{a}^{\dagger} - \alpha^* \hat{a}) },
\end{equation}
for $\alpha \in \mathbb{C} $, and where $\hat{a}$ ($\hat{a}^{\dagger}$) is the annihilation (creation) operator. Alternatively, the following description of a coherent state in the Fock basis can be derived using the normally ordered form of the operator, described in \cite{barnett2002methods}:
\begin{equation}
\label{eq:coherent state fock basis}
\ket{\alpha} = \exp{(-|\alpha|^2/2)} \sum_{n = 0}^{\infty} \frac{\alpha^n}{\sqrt{n!}}\ket{n}.
\end{equation}
In our simulation we have to truncate the Hilbert space, and therefore in order to simulate accurate states and operators we must limit the magnitude of $|\alpha|$; here we generally use a maximum value of $|\alpha| = 5$.

\subsection{Squeezed states and squeezing operators}
\label{sec:squeezing}

Squeezed states have a lower variance (in some quadrature) than coherent states, which makes them useful in a wide variety of applications, such as optical metrology \cite{caves1981quantum}. The single mode squeezed vacuum state is defined by $ \ket{\zeta}_i = \hat{S}_{i}(\zeta) \ket{0}_i $, where  $ \hat{S}_{i}(\zeta) $ is given by
\begin{equation}
\label{eq:squeezing operator fock basis}
\hat{S}_i(\zeta)=\exp{ \left[ {1 \over 2} (\zeta^* \hat{a}_i^2 - \zeta \hat{a}_i^{\dagger 2}) \right] },
\end{equation}
where $\zeta \in \mathbb{C} $ and $\hat{a}_i^\dagger$ and $\hat{a}_i$ are the creation and annihilation operators on mode $i$ \cite{barnett2002methods}. Similarly, the two mode squeezed vacuum state can be defined by 
$ \ket{\zeta}_{ij} = \hat{S}_{ij}(\zeta) \ket{0}_i\ket{0}_j $, where $ \hat{S}_{ij}(\zeta) $ is given by
\begin{equation}
\label{eq:TM squeezing operator fock basis}
\hat{S}_{ij}(\zeta)=\exp{ \left[ \zeta^* \hat{a}_i\hat{a}_j - \zeta \hat{a}_i^\dagger \hat{a}_j^\dagger \right] },
\end{equation}
where, again, $\zeta$ is a complex number which defines the state. As with coherent states, the normally ordered forms of the squeezing operators give rise to a construction of the states in the Fock basis \cite{barnett2002methods}.
 
Both the single mode and the two mode squeezed vacuum states can be created by acting on the vacuum with non-linear optical elements, such as optical parametric oscillation or four-wave mixing. They are further related because the action of a beam splitter on a two mode squeezed state is to produce two single mode squeezed states \cite{barnett2002methods}.

Squeezed states are harder to generate than, for example, coherent states, but they are still feasible and are an important resource in many experiments, so we include these in our toolbox and in most of our runs of the algorithm. We also include squeezing operators in the toolbox, although we omit these from most of our runs, as implementing a squeezing operator on an arbitrary state is difficult and rarely available in laboratories, although it is possible \cite{miwa2014exploring}.

\subsection{Beam splitters and phase shifts}
\label{sec:beamsplitters}

The phase shift operation is given by $ \exp{(i \hat{n} \phi)} $, where $\hat{n} = \hat{a}^\dagger \hat{a}$ is the photon number operator.  This is often straightforward to implement as a phase shift can be induced on a mode through changing the length of the optical path it has to travel, when compared to the other modes in the system, for example using a thermo-optic effect \cite{harris2014efficient}.

The beam splitter is a vital element of our toolbox as it allows interactions between the modes. They are also readily available, and can be implemented by a partially reflecting mirror (e.g. a half-silvered mirror) or an interface between two glued-together glass prisms, among others \cite{thorlabscatalog}. The beam splitter operation is given by
\begin{equation}
\label{eq:beam splitter operator specific}
\hat{U}_{ij} =  \exp \left[ \theta \left(\hat{a}_i\hat{a}_j^\dagger - \hat{a}_i^\dagger\hat{a}_j  \right) \right], 
\end{equation}
where the probability of a photon being transmitted through the beam splitter is $T = \cos^2{\theta}$ \cite{knott2015robust}. A `50:50' beam splitter, which mixes the two input modes in equal superposition, refers to the case where $T = 0.5$. 

\subsection{Homodyne detection}
\label{sec:homodyne}

To define homodyne detection we first require the generalised quadrature operator \cite{barnett2002methods}:
\begin{equation}
\label{eq:general quadrature operator}
\hat{x}_\lambda = \frac{1}{\sqrt{2}} \left[ \hat{a} \exp{(-i\lambda)} + \hat{a}^\dagger \exp{(i\lambda)} \right],
\end{equation}
where $\lambda$ is the quadrature angle. Setting $\lambda = 0$ and $\lambda = \pi/2$ in the above expression gives the usual `position' and `momentum' quadrature operators, respectively. Homodyne detection is a projection onto the eigenstates of these quadrature operators. However, it is not possible to construct normalised eigenstates of the quadrature operators. Instead we construct states, $\ket{x_\lambda}$, defined by a complex number $x_\lambda = |x_\lambda|\exp{i\lambda}$, which obey the eigenvalue equation
\begin{equation}
\label{eq:quadrature operator eigenvalue equation}
\hat{x}_\lambda \ket{x_\lambda} = |x_\lambda| \ket{x_\lambda}.
\end{equation}
These are not orthogonal but instead their overlap is given by the Dirac delta function
\begin{equation}
\label{eq:quadrature eigenstate overlap}
\braket{x_\lambda | x'_\lambda} = \delta \left(x_\lambda - x'_\lambda \right).
\end{equation}
These eigenstates may also be constructed by operating on the vacuum state as follows \cite{barnett2002methods}
\begin{equation}
\label{eq:quadrature eigenstate construction}
\ket{x_\lambda} = \pi^{1/4} \exp \left[-\frac{1}{2} |x_\lambda|^2 + \sqrt{2} x_\lambda \hat{a}^\dagger - \frac{1}{2} \exp(2i\lambda) \hat{a}^{\dagger 2}  \right] \ket{0}. \notag
\end{equation}

Perfect homodyne detection is the projection onto a line in phase space, characterised by the projector $\ket{x_\lambda}\bra{x_\lambda}$.

\subsection{Simulating noisy experiments}
\label{APX:noise}

Photon loss can be modelled by mixing the noisy mode with the vacuum using a beam splitter of transmissivity $T = 1 - \gamma$ (and hence loss rate $\gamma$) \cite{knott2015robust}. However, this is computationally expensive as it requires simulating an additional mode, so here we simulate loss using Kraus operators as follows \cite{nielsen2002quantum}
\begin{equation}
\label{eq:Kraus operators}
\rho_{out} = \sum_{k=0}^\infty K_k \ket{\psi_{in}}\bra{\psi_{in}} K_k^\dagger,
\end{equation}
where $\rho_{out}$ is the output state, $\ket{\psi_{in}}$ is the input state, and
\begin{subequations}
\label{eq:Kraus ops for loss}
\begin{align}
K_k = \sum_{n=0}^\infty \sqrt{\binom{n}{k}}\sqrt{(1-\gamma)^{n-k}\gamma^k}\ket{n-k}\bra{n}.
\end{align}
\end{subequations}
The Kraus operator $K_k$ represents the loss of $k$ photons to the environment with loss rate of $\gamma$.

We can include loss in the measurements by modifying the POVM elements to \cite{scully1969quantum, kok2007linear}
\begin{equation}
\label{eq:number measurement POVM}
E_n = \sum_{k=n}^\infty \binom{k}{n}(1-\gamma)^n\gamma^{k-n}\ket{k}\bra{k}. 
\end{equation}
For the bucket measurement, we use \cite{kok2000postselected, kok2007linear}
\begin{subequations}
\label{eq:bucket measurement POVM}
\begin{align}
E_0 &= \sum_{n=0}^\infty \gamma^n\ket{n}\bra{n}, \\
E_1 &= \mathbb{I} - E_0,
\end{align}
\end{subequations}
where $E_0$ corresponds to a measurement of no photons and $E_1$ is the measurement of one or more photons.
Finally, for the multiplex detector, we construct a set of POVMs by using Eq.~(\ref{eq:multiplex POVM}) but with $|c\rangle\langle c|$ replaced with $E_{n=c}$ from Eq.~(\ref{eq:number measurement POVM}).

\section{Genetic algorithms}
\label{APX:ga_details}

Genetic algorithms take inspiration from biological evolution \cite{matlabga,goldberg1989, conn1991globally}. The aim of the algorithm is to maximise (or minimise) the fitness function -- a summary of how this is achieved is presented in Fig.~\ref{fig:geneticalgorithmflowchart}.  Before starting the algorithm, the fitness function must be defined. This is the function to be optimised, known as the objective function in standard optimisation algorithms, which must produce a real number (the \emph{fitness} or \emph{objective value}) from a collection of real variables. Any point that the fitness function can be applied to, i.e. a list of values for each of the variables, is known as a \emph{genome}, and the individual values of variables in this genome are known as the genes.

The algorithm works by varying a group of points, known as the population. Hence, the first step is to generate the initial population. For each genome in the population, each variable is generated at random within the provided bounds. Then the algorithm must iterate the population, hoping to optimise the fitness function by finding genomes that improve on the best value of the fitness function of previous generations. This iteration has the following steps, which repeat until the stopping conditions are reached:
\begin{description}
	\item [Evaluation] The value of the fitness function is found for each genome in the current population, these are known as the raw fitness scores. At this point the stopping conditions are also evaluated.
	\item [Selection] Using the results of the evaluation, a subset of the population known as the parents are selected.
	\item [Reproduction] The next generation is generated using the selected parents. This is done using three methods: elite, crossover and mutation.
\end{description}

The selection step works by first scaling the raw fitness scores to convert them into a more usable range of values. These scaled scores are then used to select the parents of the next generation. The selection function assigns a higher probability of selection to genomes with higher scaled fitness scores. This mimics natural selection, or `survival of the fittest', as the `fitter' genomes are more likely to reproduce.

The reproduction step creates genomes of the next generation (`children') using the parents of the current generation through three methods. The first are elite children, which form only a small percentage of the next generation. These are the parents with the highest fitness scores, which survive to the next generation unchanged. The second are crossover children. Here, two parents are chosen at random and each gene of the child is produced by copying the gene from one of the two parents. The final method to create children is mutation. Here, a single parent is chosen and random changes are made to its genome to produce the child. In Section \ref{hyperparameters_settings} we introduce the specific mutation and crossover functions we use.

These children together form the next generation and the process repeats until the stopping conditions are met (see \cite{legos_blog} for a visual introduction to genetic algorithms). Possible stopping conditions include meeting a maximum number of generations over which the best fitness score does not change, within tolerance. There is no proof that genetic algorithms must converge \cite{matlabga}, but they have been known to perform well in situations where standard, gradient-based, optimisation algorithms have failed. In quantum circuit design, genetic algorithms have been successfully used to design quantum logic gates \cite{barnum2000quantum}. In other fields, examples of problems where genetic algorithms have been used include a NASA design of a radio antenna to pick up signals in space \cite{hornby2006automated}, computer models for walking for bipedal creatures \cite{geijtenbeek2013flexible} and optimising the aerodynamics of hypersonic space vehicles \cite{evans2017aerodynamic}.

\section{Running AdaQuantum to obtain our results}
\label{APX:running_for_results}

\subsection{Parameters that affect the running speed}

In general it is not easy to say \textit{quantitatively} how the speed and performance of the algorithm changes when changing different parameters, such as the number of modes or the size of the selected toolbox (which in turn determine the genome length). The reason for this is that the dominant contribution to the run-time is the truncation. But, as discussed in Section \ref{sec:three_stage}, in Stage 3 of our algorithm the truncation changes with each simulation.

With this in mind, increasing the number of allowed operators (labeled $m$ in Fig.~\ref{fig:stateengineeringscheme}) in itself would (approximately) linearly increase the run-time. But in addition, having more operators allows for the creation of larger states (e.g. by applying multiple displacement operators), so the point where we truncate in general needs to increase to accommodate the higher-energy states that will be produced. For the most of the results in this paper, we allowed for $m=3$ operators. We also ran the algorithm for $m=5$ for some of the no-loss runs, but didn't see any improvement. One important point is that we included the \textit{identity} in all runs of the algorithm, which explains why many of the results in this paper only have 2 operators. Another dominant factor in the simulation speed is the toolbox selection, particularly in the choices of parameters. Allowing for large states and operators that increase the energy slows down the simulation significantly. However, in our runs we always limited the size of the output state to either 1 or 2 photons on average (this can be done in the user interface for selecting the fitness function), which means that states that are too large \textit{after} the measurement are discarded anyway.

All the runs of AdaQuantum presented here are for two-mode experiments. We attempted three-mode runs but were unable to outperform the two-mode experiments. The reason for this is that the simulation becomes exponentially slower when we increase the number of modes, and furthermore the search space becomes much larger. As elaborated on in Section \ref{sec:discussion}, improving the genetic algorithm so that it can effectively search for three-mode experiments will be a central focus of our future work.

\begin{figure*}
\includegraphics[trim=0cm 0cm 0cm 1.8cm, clip=true, width=18cm]{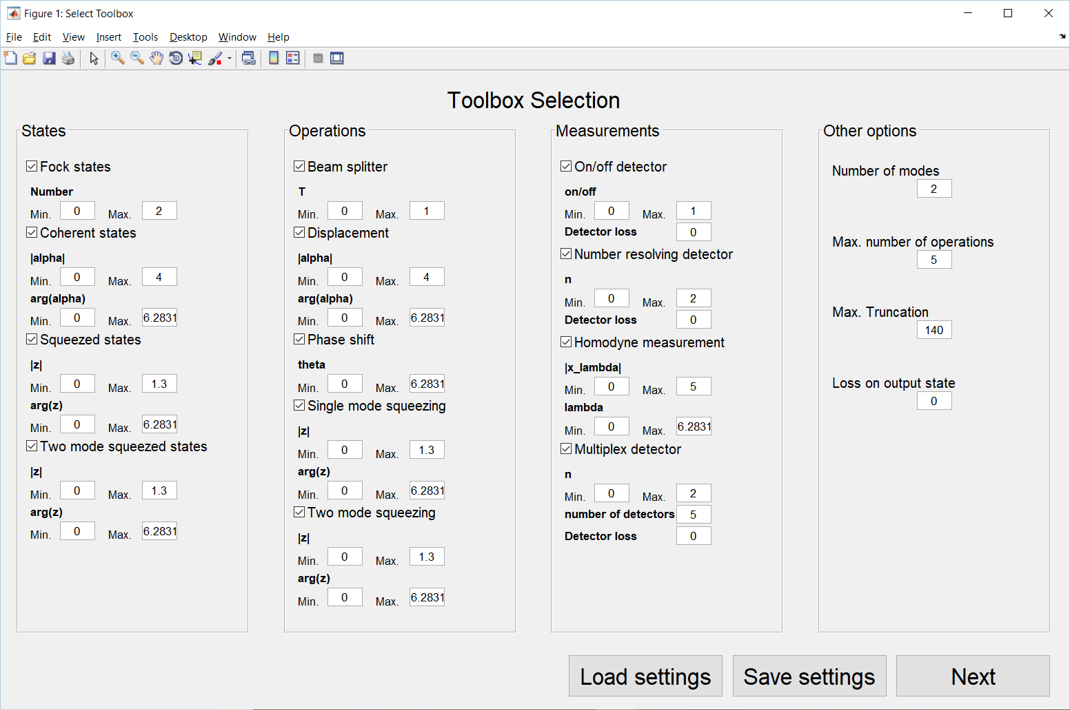}\\
	\caption{The user interface that opens when first running our program, allowing the user to choose the elements of the toolbox they have available, and the parameter limit for these elements. After pressing ``Next'', the user is taken to a similar user interface that allows the fitness function to be specified. After this has been chosen the genetic algorithm runs and generates optimised experiment designs.}
\label{fig:UI - toolbox}
\end{figure*}

\subsection{Hyperparameters and settings for the genetic algorithm}
\label{hyperparameters_settings}

Mutation and crossover are central to genetic algorithms, and the choices of mutation and crossover functions, as well as the hyperparameters of these functions, plays a large part in determining the effectiveness of the search. In our runs we experimented with three different crossover functions that are all inbuilt into Matlab \cite{matlabga}: \texttt{scattered}, \texttt{single point} and \texttt{two point} crossover. None of the crossover functions exploit the division into states, operators and measurements, for example by mixing the input states for one experiment with the operators and measurements of another. It would be interesting to see whether exploiting such a division improves the results. In general, it isn't intuitively clear why crossover works, given the nature of quantum optics experiments, but our tests demonstrated that it was an important element of the search process, increasing diversity and enabling a larger space to be explored. Despite this, we suspect that more in-depth tests (that we reserve for future work) might reveal that the advantage of crossover comes more from the occasions when it mixes similar experiments with each other. In the next subsection we give the specific values of the genetic algorithm settings and hyperparameters for the results presented in this paper.


The mutation functions we used mutated both the experiment arrangements and the parameters, and therefore our search is performed simultaneously over these quite diverse characteristics of an experiment. Matlab did not have a mutation function with enough flexibility for our purposes, so we introduced two mutation functions, \texttt{power mutation} and \texttt{power selection}. Both are based on \cite{deep2007new}. In short, \texttt{power mutation} mutates every gene in the genome by a random distance, whose maximum magnitude is determined by the value of a hyperparameter named \texttt{power}, for which \texttt{power}$ =1$ will mutate each gene to a completely random new value, whereas \texttt{power}$ =\infty$ doesn't mutate at all.  \texttt{power selection} only mutates a fraction (given by \texttt{rate}) of the genes. See \cite{AdaQuantum_GitHub} and \cite{deep2007new} for further details.


Throughout we used tournament selection to select the parents for the next generation.

\subsection{Obtaining the results in this paper}

We mainly optimised our search algorithm using the toolbox labelled `Tool 1' in Fig.~\ref{fig:Fovernbar_vs_nbar} and Table.~\ref{results_noloss_table}. The main reason for this being that this allowed us to compare directly with the results in \cite{knott2016search}. For our initial runs, using Matlab's inbuilt genetic algorithm with a population size of 200, the search often converged to the same results in \cite{knott2016search}. But after increasing the population size, using our 3-stage algorithm, and optimising the hyperparameters, we soon found the greatly-improved results presented in this paper. Having found a variety of genetic algorithm settings and hyperparameters, we then ran AdaQuantum for all the toolboxes presented in this paper. To obtain each result, we ran the AdaQuantum for $96$ hours on $16$ cores of the University of Nottingham's High Performance Computing facility. For most of the toolboxes and fitness functions, we ran the algorithm twice with different settings/hyperparameters, then selected the better of the two results. The settings and hyperparameters are shown in Table.~\ref{settings_hyperparameters}. The population sizes in stages 1, 2 and 3 were $10^7$, $10^5$ and $2$x$10^4$, respectively.

\begin{table*} [t]
{\renewcommand{\arraystretch}{1.5} 
\begin{tabular}{lllllll}
FF \& Toolbox ~& crossover function ~& crossover fraction ~& tournament size ~& mutation function ~& \texttt{power} ~& \texttt{rate} \\
\hline
Pure QFI \& Tool 1 ~~~ & \texttt{scattered} & 0.3 & 8 & \texttt{power mutation} & 10 & -\\ 
Pure QFI \& Tool 2 & \texttt{single point} ~~~ & 0.3 & 8 & \texttt{power selection} ~~~ & 4 & 0.2\\ 
Pure QFI \& Full & \texttt{two point} & 0.3 & 8 & \texttt{power mutation} & 10 & -\\ 
Pure QFI \& No PNRD ~~& \texttt{two point} & 0.3 & 8 & \texttt{power mutation} & 10 & -\\ 
~ & ~ & ~ & ~ & ~ & ~ & ~ \\
QFI, loss = 0.05 & \texttt{single point} & 0.3 & 8 & \texttt{power selection} & 10 & 0.5\\ 
QFI, loss = 0.1 & \texttt{two point} & 0.3 & 8 & \texttt{power selection} & 20 & 0.1\\ 
QFI, loss = 0.2 & \texttt{scattered} & 0.3 & 8 & \texttt{power mutation} & 5 & -\\ 
QFI, loss = 0.3 & \texttt{scattered} & 0.3 & 8 & \texttt{power mutation} & 5 & -\\ 
~ & ~ & ~ & ~ & ~ & ~ & ~ \\
BMSE, $\mu = 1$ & \texttt{scattered} & 0.3 & 8 & \texttt{power mutation} & 10 & -\\ 
	\end{tabular}}
\caption{The settings and hyperparameters for the results in this paper. FF = fitness function. For all runs the number of elite children was 10. Despite the crossover rate being $0.3$ for all results here, we found that values between $0.2$ and $0.5$ often gave comparable results, suggesting the crossover does provide some benefit. Unfortunately we do not have available the hyperparameters for the BMSE runs for the remaining values of $\mu$. However, we found that for the BMSE the search converged quickly for all the hyperparameters.}
\label{settings_hyperparameters}
\end{table*}

It is important to stress here that the focus of this paper is not on optimising the genetic algorithm. Rather, our goal was to create a search algorithm that worked reliably for a number of different fitness functions, and to find a range of hyperparameters and settings such that the algorithm works consistently. Researchers who use AdaQuantum with a different fitness function will likely have to experiment with different settings/hyperparameters, but our choices in Table.~\ref{settings_hyperparameters} should serve as a useful guide. In ongoing work we are performing a thorough optimisation of the settings/hyperparameters using a range of toolboxes and fitness functions, and comparing this to a range of different search algorithms.

We made a number of observations when running the algorithm. Some of the runs for pure states (in particular for the larger toolboxes) did not fully converge after the 96 hours time, suggesting that improvements over the results here might be possible. Furthermore, for most of the pure-QFI fitness functions, different runs of the algorithm often produced different experiments with similar fitness values, suggesting there are whole classes of states with large QFI values. In contrast, for the mixed state runs and the BMSE the algorithm often converged quickly (long before the 96 hour running time) to the same or similar experiments, even for a range of hyperparameters/settings.

\section{Comparison to previous work}
\label{APX:comparison}

Ref.~\cite{knott2016search} is most similar to our own, as it too uses a search algorithm to optimise the arrangement and parameters in a quantum optics setup to find quantum states for metrology. However, our current algorithm has numerous significant advantages over \cite{knott2016search}: i) The search algorithm in \cite{knott2016search} is less refined, and is in effect a stochastic hill-climbing algorithm with random restart. ii) The simulation in our current algorithm has overcome various numerical challenges present in \cite{knott2016search}, which meant that the search algorithm in \cite{knott2016search} had to truncate the number basis at 30 photons, in comparison to our current work that allows the truncation to reach 170. The combined effect of the developments i) and ii) is that the results we present here greatly exceed those in \cite{knott2016search}, with up to a 5-fold QFI improvement (and note that truncating at 30 photons does not even allow any of the states we found here to be properly simulated).

Further advantages over \cite{knott2016search} include: iii) In our current work we model the dominant experimental noise in typical quantum optics experiments (photon loss on the output state and imperfect detectors). iv) The algorithm we present here is available on GitHub \cite{AdaQuantum_GitHub}, and as we detail in Section \ref{sec:flexibility} it has been designed specifically for flexibility (e.g. AdaQuantum can optimise with different numbers of modes and toolboxes), as illustrated by the three different fitness functions we study here (compared to just the QFI in \cite{knott2016search}). AdaQuantum can also be easily modified, so that researchers can add/change both the quantum optics elements to be searched over, and the fitness function to be optimised. A long term goal of our research is to construct an algorithm that can fully model, and then optimise and design, realistic and general quantum optics experiments. Together iii) and iv) mean that AdaQuantum has taken a major step towards this goal.

Refs.~\cite{krenn2016automated} and \cite{melnikov2018active} differ from ours in both the optics experiments of interest (they search for higher-dimensional states with angular momentum, and look for certain entangled states) and the algorithms used (\cite{krenn2016automated} uses random search with learning, and \cite{melnikov2018active} uses reinforcement learning). At present the optics settings are different enough so that our algorithm cannot be directly applied to their research, and vice versa, though there is no fundamental reason why future work cannot apply global search algorithms to their setting, and reinforcement learning to ours.

Finally, in Refs.~\cite{arrazola2018machine} and \cite{sabapathy2018near} a machine learning algorithm can be used to optimise the parameters in an optics experiment. \cite{arrazola2018machine,sabapathy2018near} have the advantage that they use gradient decent to learn the parameters, rather than our stochastic algorithm, but this seems to come at the expense of the speed of their simulation (we have not done a full comparison, but the online version of the algorithm they use, \texttt{Strawberry Fields} \cite{killoran2018strawberry}, has to truncate the Fock basis at 20 photons). Furthermore, we optimise over the arrangement of elements in addition to the parameters; our algorithm is specifically designed so that the fitness function can be easily modified; and we have realistic experimental setups in mind (though we believe these latter two features could be incorporated into \cite{killoran2018strawberry}).

\end{widetext}

\end{document}